%% file: bhh.tex
      [1999/12/01 v1.4c Il Nuovo Cimento]
\documentclass{cimento}

\usepackage[abs]{overpic}
\usepackage{graphicx}  
\title{Observation of New Charmless Decays of Bottom Hadrons}
\author{Michael J.~Morello\from{ins:x}\thanks{{\tt morello@fnal.gov}}
}
\instlist{\inst{ins:x}Fermi National Accelerator Laboratory
}

\PACSes{\PACSit{13.25.Hw}{Decays of bottom mesons}
\PACSit{13.30.Eg}{Hadronic decays}}

\include{macro}

\begin{document}

\maketitle

\begin{abstract}
We search for new charmless decays of neutral $b$--hadrons to pairs of 
charged hadrons with the upgraded Collider Detector at the Fermilab Tevatron.
Using a data sample corresponding to \mbox{1\lumifb} of integrated luminosity, 
we report the first observation of the \BsKpi\ decay, with a significance of 
$8.2\sigma$, and measure 
$\BR(\BsKpi)= (5.0 \pm 0.7\stat \pm 0.8\syst)\times 10^{-6}$. 
We also report the first observation of charmless $b$--baryon decays
in the channels \Lbppi\ and \LbpK\ with significances
of $6.0\sigma$ and $11.5\sigma$ respectively, and we measure 
$\BR(\Lbppi) = (3.5 \pm 0.6\stat \pm 0.9\syst)\times 10^{-6}$ and 
$\BR(\LbpK) = (5.6 \pm 0.8\stat \pm 1.5\syst)\times 10^{-6}$. 
No evidence is found for the decays \BdKK\ and \Bspipi, and we set an 
improved upper limit $\BR(\Bspipi) < 1.2\times 10^{-6}$ at the 90\% confidence level.
All quoted branching fractions are measured using $\BR(\BdKpi)$ as a 
reference.
\end{abstract}

\section{Introduction}

Non-leptonic two-body charmless decays of neutral $b$ hadrons 
(\Bdhh, \Bshh\ and \Lbph, where $h$ is a charged pion or kaon)
are very interesting for the understanding
of flavor physics and \CP\ violation mechanism in the $b$-hadron sector. 
Their rich phenomenology offers several opportunities to explore and constrain
the parameters of the quark-mixing matrix (\ie\ Cabibbo-Kobayashi-Maskawa, CKM).
These processes allow to access the phase of the $V_{ub}$ element of the 
CKM matrix ($\gamma$ angle), and to test the reliability of the Standard Model (SM) and hadronic calculations.
The presence of New Physics can be revealed by its impact on 
their decay amplitudes, where new particles may enter in penguin diagrams.
The \BdKpi\ is the first process involving the $b$ quark
where direct CP violation has been observed.

The measurements obtained at $e^{+}e^{-}$ 
colliders (ARGUS, CLEO, LEP, and more recently, BaBar and Belle experiments) already 
provided a wealth of results for \Bd\ and \Bu\ mesons.  
The upgraded Collider Detector at the Fermilab Tevatron (CDF~II), with its large production of 
$b$-hadrons is in principle an ideal environment for studying these rare modes.
In addition to providing further large samples of \Bd\ and \Bu\ mesons in a different experimental 
environment, it provides the exciting opportunity of studying the charmless decays of 
other $b$-hadrons that are unaccessible (or much less accessible) in other experiments.
A variety of techniques have been proposed to constrain the CKM parameters 
or probe effects of New Physics \cite{fleischer,th:uspin_fit,gronau_uspin,Lipkin:2005pb} 
exploiting a combination of observables from \Bs\ and \Bd\ , \Bu\ mesons.

The \BsKpi\ decay offers several and interesting strategies to extract 
useful information from the comparison between its observables and those of its U-spin related 
partner \BdKpi~\cite{gronau_uspin,Lipkin:2005pb}.
By combining the information  of rates and direct \CP\ asymmetries of U-spin--related decays \BdKpi\ and 
\BsKpi\ \cite{gronau_uspin,Lipkin:2005pb}  allows a stringent test of the Standard Model 
origin of the $\mathcal{O}(10\%)$  direct CP asymmetry observed in \BdKpi\ \cite{babelle_acpbdkpi},  
which is not matched by a similar 
effect in the $B^{+}\to K^{+}\pi^{0}$ decay, which differs only by the spectator quark. 
This raised discussions about a possible exotic source for the \CP\ violation in the \BdKpi\ decay 
\cite{th:acp_Bd_Bu_sanda,th:acp_Bd_Bu_beneke,gronau_acp}.
Any significant disagreement
between the measured partial rate asymmetries of strange and non-strange $b$-meson $K\pi$ decays 
should be strong indication of  New Physics. 

The \BsKpi\ is still unobserved and the current experimental 
upper limit $\BR(\BsKpi)<5.6\times10^{-6} ~@~90\%~{\rm CL}$ \cite{Abulencia:2006psa} from CDF is 
very close to (sometimes lower than) the current theoretical expectations \cite{th:QCDf,bhh_pQCD,bskpi_yu,bhh_scet}.
 The comparison of this branching fraction, sensitive to CKM angle values of  $\alpha$ and $\gamma$ \cite{bskpi_yu},
with theoretical predictions provides valuable information for tuning the 
phenomenological models of hadronic $B^0_{(s)}$ decays and for optimizing the choice of their input parameters. 
Therefore, in this context the measurement of the decay rate of the \BsKpi\ and the measurement 
of its direct CP asymmetry becomes crucial.

The amplitudes of penguin-annihilation and exchange diagrams, in which 
all initial-state quarks undergo a transition, are difficult to predict with current phenomenological 
models. In general they may carry different \CP--violating and \CP--conserving phases with respect to the 
leading processes, thereby influencing the determination of CKM-related parameters. 
The \BdKK\ and \Bspipi\ decays proceed only through these kinds of diagrams. A  simultaneous 
measurement of their decay rates 
(or improved constraints on them) would provide valuable estimates of the magnitude of these 
contributions \cite{buras_npb}.

Simultaneous measurements of \Bhh\ observables, in most cases, exploit the U-spin 
symmetries to partially cancel out or constrain hadronic uncertainties
and probe the electroweak and QCD structure. 
U-spin symmetry is not exactly conserved in the Standard Model and the 
magnitude of its violation is not precisely known but 
most authors estimate a $\mathcal{O}(10\%)$ effect. 
The \Bhh\ system  is a privileged laboratory since it offers the 
simultaneous opportunities of using U-spin assumptions and, at the same time, 
 of checking their validity by measuring the symmetry breaking-size,  
from the interplaying of several U-spin--related observables. 

Two--body charmless decays are also expected from bottom baryons. 
The modes \LbpK\ and \Lbppi\ are predicted to have measurable branching 
fractions, of order $10^{-6}$~\cite{Mohanta:2000nk}, and, in
addition to the interest in their observation, must be considered
as a possible background to the rare \Bs\ and \Bd\ modes being investigated.

In this Letter we report the results of a
search for rare decays of neutral bottom hadrons into a pair of charged
charmless hadrons ($p$, $K$ or $\pi$), performed in 1\lumifb\ of \ppbar\ 
collisions at $\sqrt{s} = 1.96$ TeV, collected 
by the upgraded Collider Detector (CDF II) at the Fermilab Tevatron.
We report the first observation of modes \BsKpi, \LbpK, and 
\Lbppi, and measure their relative branching fractions. 
This is a short overview of the work documented in Ref.~\cite{my_thesis} and 
the results were published in~\cite{Aaltonen:2008hg}.

Throughout this paper, C-conjugate modes
are implied and branching fractions indicate
\CP-averages unless otherwise stated.

\section{\cdfii\ detector}

The \cdfii\ detector~\cite{CDF_det,runIITDR}, in operation since 2001, 
is an azimuthally and forward-backward symmetric apparatus designed to 
study $p\bar{p}$ collisions at the Tevatron. It is
a general purpose solenoidal detector which 
combines precision charged particle tracking with fast projective calorimetry 
and fine grained muon detection.
Tracking systems are contained in a superconducting solenoid, 1.5~m in radius 
and 4.8~m in length, which generates a 1.4~T magnetic field parallel to the 
beam axis. Calorimetry and muon systems are all outside the solenoid.
The main features of the detector systems are summarized below.

The tracking system consists of a silicon microstrip system~\cite{runIISi} and of 
an open-cell wire drift chamber~\cite{runIICOT} that surrounds the silicon.
The silicon microstrip detector consists of seven layers (eight layers for
$1.0<|\eta|< 2.0$) in a barrel geometry that extends from a radius of 
$r = 1.5$ cm from the beam line to $r = 28$ cm. The layer closest to the beam pipe
is a radiation-hard, single sided detector called Layer\O\O\ which employs         
sensors supporting high-bias voltages. This enables signal-to-noise
performance even after extreme radiation doses. The remaining seven layers are
radiation-hard, double sided detectors. 
The first five layers after Layer\O\O\ comprise 
the Silicon VerteX II (SVXII) detector and the two outer layers comprise
 the Intermediate Silicon Layer (ISL) system. This entire system 
allows track reconstruction in three dimensions. 
The impact parameter resolution of the combination
of SVXII and ISL is about 48~$\mu$m including a 30~$\mu$m contribution from the beamline for tracks 
with transverse momentum of 2~\pgev.
The $z_0$ resolution of the SVXII and ISL is 70~$\mu$m.
The 3.1~m long cylindrical drift chamber (COT) covers the radial range from 
40 to 137 cm and provides 96 measurement layers, organized 
into alternating axial and $\pm 2^{\circ}$ stereo superlayers.
The COT provides coverage for $|\eta| \leq $1. The hit position resolution is 
approximately 140 $\mu$m and the momentum resolution 
$\sigma_{p_{T}}/p_{T} \simeq 0.15\%\, p_{T}$/(GeV/$c$). 
This corresponds to an observed mass-widths of about 14~\massmev\
for the $J/\psi\to\mu^+\mu^-$ decays, and of about 8 \massmev\ for the \Dkpi\ decays.
The specific energy loss by ionization (\dedx) of charged particles in the COT can
be measured from the amount of charge collected by each wire.
This yields a nearly-constant separation of 1.5 standard deviations
between pions and kaons over the range $2<p_{T}<10$~GeV/$c$.

A Time-of-Flight (TOF) detector~\cite{runIITOF}, based on plastic scintillators and 
fine-mesh photomultipliers is installed in a few centimeters clearance just outside the
COT. The TOF resolution is $\approx 100$ ps and it provides at least two standard 
deviation
separation between $K^{\pm}$ and $\pi^{\pm}$ for momenta $p <$ 1.6 GeV/c. 

Segmented electromagnetic and hadronic sampling calorimeters surround the
tracking system and measure the energy flow of interacting particles in the 
pseudo-rapidity range $|\eta|<$ 3.64.
The central calorimeters (and the endwall hadronic calorimeter) cover the 
pseudorapidity range $|\eta|<$ 1.1(1.3). 
The central electromagnetic calorimeter~\cite{runIICEM} (CEM) uses lead sheets 
interspersed with polystyrene
scintillator as the active medium and employs phototube readout.
Its energy resolution is $13.5\%/\sqrt{E_T} \oplus 2\%$.
The central hadronic calorimeter~\cite{runIICHA} (CHA) uses steel absorber 
interspersed with acrylic scintillator as the active medium.
Its energy resolution is $75\%/\sqrt{E_T} \oplus 3\%$.
The plug calorimeters cover the pseudorapidity region  1.1 $<|\eta|<$ 3.64.
They are sampling scintillator calorimeters which are read out with plastic 
fibers and phototubes. The energy resolution of the plug electromagnetic 
calorimeter~\cite{runIIPEM} is $16\%/\sqrt{E} \oplus 1\%$.
 The energy resolution of the plug hadronic 
calorimeter is $74\%/\sqrt{E} \oplus 4\%$. 

The muon system resides beyond the calorimetry. Four layers of planar drift chambers (CMU)
detect muons with $p_T >$ 1.4 GeV/c which penetrate the five absorption lengths of
calorimeter steel. An additional four layers of planar drift chambers (CMP) instrument
0.6~m of steel outside the magnet return yoke and detect muons with $p_T >$ 2.0 GeV/c. 
The CMU and CMP chambers each provide coverage in the pseudo-rapidity range 
$|\eta|<0.6$. The Intermediate MUon detectors (IMU) are covering the region 
1.0 $<|\eta|<$1.5. 

The beam luminosity is determined by using gas Cherenkov counters located in 
the $3.7 < |\eta| < 4.7$ region which measure the average number of inelastic
$p\bar{p}$ collisions per bunch crossing~\cite{runIIlum}.

The trigger and data acquisition systems are designed to accommodate  the
high rates and large data volume of Run II. Based on preliminary information  
from tracking, calorimetry, and muon systems, the output of the first  level of
the trigger is used to limit the rate for accepted events  to $\approx$
18~kHz at the luminosity range of 3-7 10$^{31}~{\rm cm}^{-2}{\rm s}^{-1}$. 
At the next trigger stage, with more refined information and additional tracking
information from the silicon detector, the
rate is reduced further to $\approx$ 300~Hz.  The third and final level of
the trigger, with access to the complete event information, uses software
algorithms and a computing farm, and reduces the output rate to $\approx$
75~Hz, which is written to permanent storage.

The only physics objects used in this analysis are the tracks, then just tracking system 
has been used. 

\section{\label{sec:Data}Data sample}

We analysed an integrated luminosity  $\int\Lumi dt\simeq 1$~\lumifb\ sample of pairs of oppositely-charged particles
with $p_{T} > 2$~\pgev\ and   $p_{T1} + p_{T2} > 5.5$~\pgev,
used to form $b$-hadron candidates.
The trigger required also a transverse opening angle $20^\circ < \Delta\phi < 135^\circ$ between the two tracks,
  to reject background from particle 
pairs within the same jet and from back-to-back jets.
In addition, both charged particles were required to originate from
a displaced vertex with a large impact parameter $d$ (100 $\mu$m $< d < 1$~mm), 
while the $b$-hadron candidate was required to be produced in
the primary $p\bar{p}$ interaction ($d_{B}< 140$~$\mu$m) and to have travelled a transverse distance
$L_{T}>200$~$\mu$m.

The offline selection is based on a more accurate 
determination of the same quantities used in the trigger, with the 
addition of two further observables: 
the isolation ($I_{B}$) of the candidate~\cite{Isolation},
and the quality of the three-dimensional fit ($\chi^{2}$ with 1 d.o.f.) of the 
decay vertex of candidate. 
Requiring a large value of $I_{B}$ reduces the background from 
light-quark jets, and a low $\chi^{2}$ reduces the 
background from decays of different long-lived particles within the event, 
owing to the good resolution of the SVX detector in the $z$ direction.

In the offline analysis, an unbiased optimization procedure determined a
 tightened selection on track-pairs fit to a common decay vertex.
We chose the selection cuts minimizing directly the expected uncertainty of the physics
observables to be measured. 
The selection is optimized for detection of the \BsKpi\ mode.
Maximal sensitivity for both discovery and limit setting
is achieved with a single choice of selection requirements~\cite{gp0308063} 
by  minimizing the variance of the estimate of the branching  fraction
in the absence of signal~\cite{my_thesis}.
The variance is evaluated by performing the full measurement
procedure on simulated samples containing background and all signals  from
the known modes, but no \BsKpi\ signal.
The variance has been parameterized with analytical functions of the signal yield ($\mathsf{S}$)
and background level ($\mathsf{B}$), and the free parameters determined from analysis of pseudo-experiments
reproducing the experimental circumstance of data.
For each set of cuts, $\mathsf{S}$ was estimated from Monte Carlo
simulation and normalized to the yield observed in data after the
trigger selection, and $\mathsf{B}$ was extrapolated from the sidebands of the $\pi\pi$-mass distribution in data.
This procedure yields the final selection: $I_B>0.525$, $\chi^{2}<5$,
$d > 120~\mu$m, $d_{B} < 60~\mu$m, and $L_{T} > 350~\mu$m.

\begin{figure}[htb]
\center
\begin{overpic}[scale=0.33]{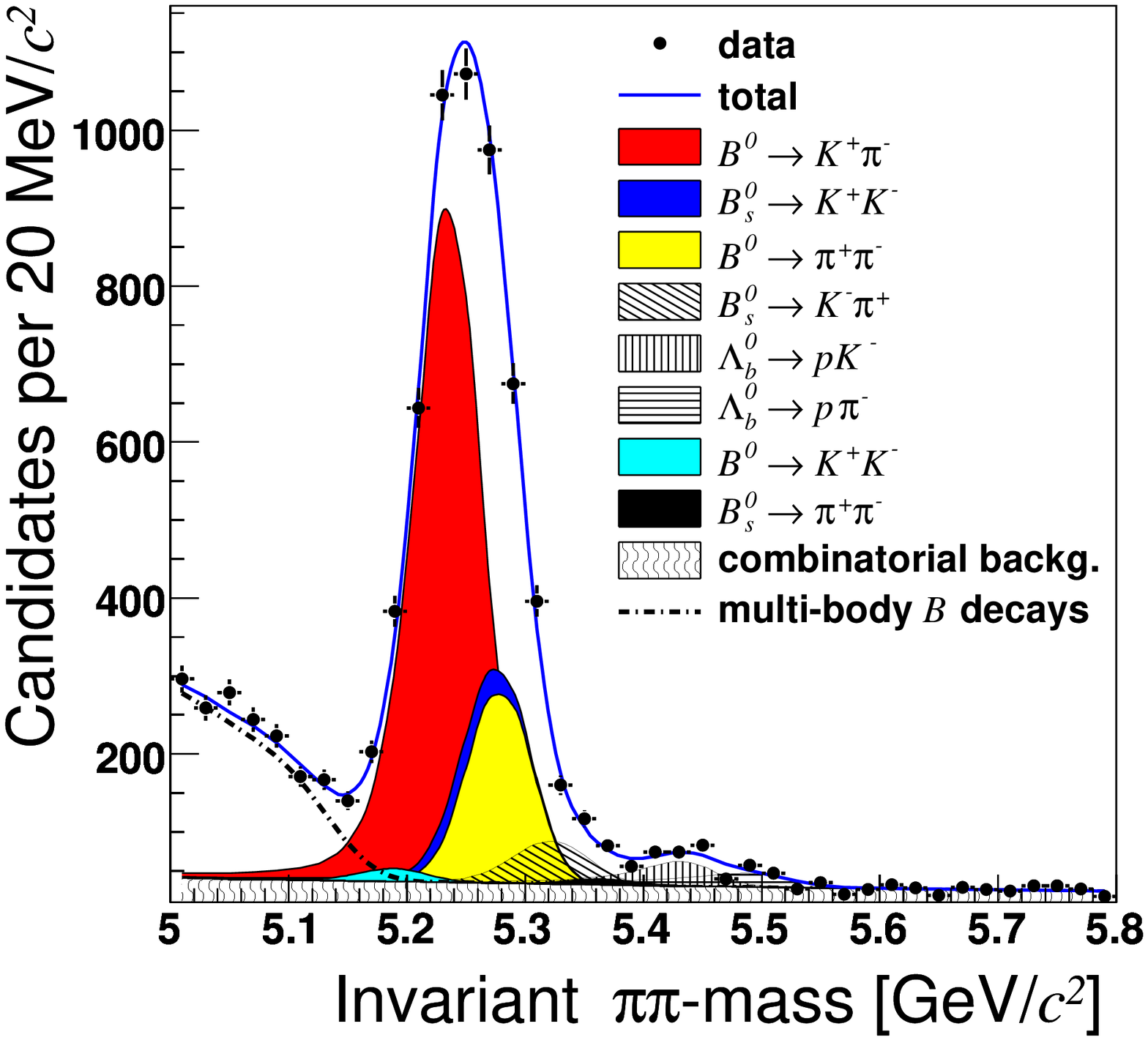}
\put(160,150){(a)}
\end{overpic} 
\begin{overpic}[scale=0.33]{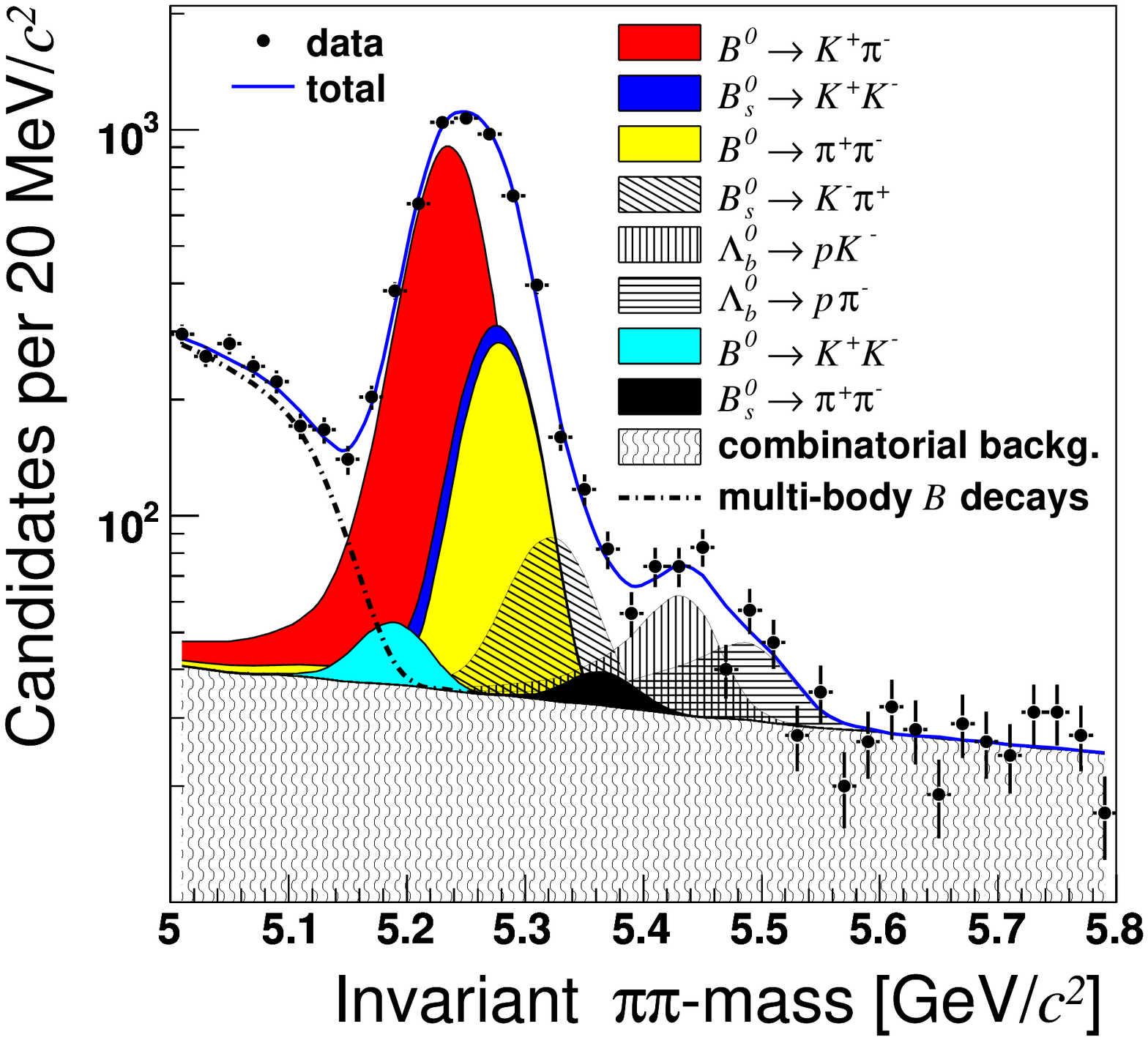}
\put(160,150){(b)}
\end{overpic}   
\caption{Invariant $\pi\pi$-mass distribution
of reconstructed candidates. The charged pion mass is assigned to both tracks.
The total projection and projections of each signal and background 
component of the likelihood fit are overlaid on 
the data distribution. Signals and multi-body $B$ background components are shown stacked on the combinatorial 
background component. Linear scale (a), logarithmic scale (b).
}
\label{fig:projections}
\end{figure}

The resulting invariant-$\pi\pi$-mass distribution (see Fig.~\ref{fig:projections}) shows a clean signal of \bhh\ decays.
In spite of a good mass resolution ($\approx 22\,\massmev$), the various signal decay modes overlap into an unresolved
mass peak.

\section{\label{sec:fit}Fit of composition}
\begin{figure}[htb]
\begin{center}
\includegraphics[scale=0.22]{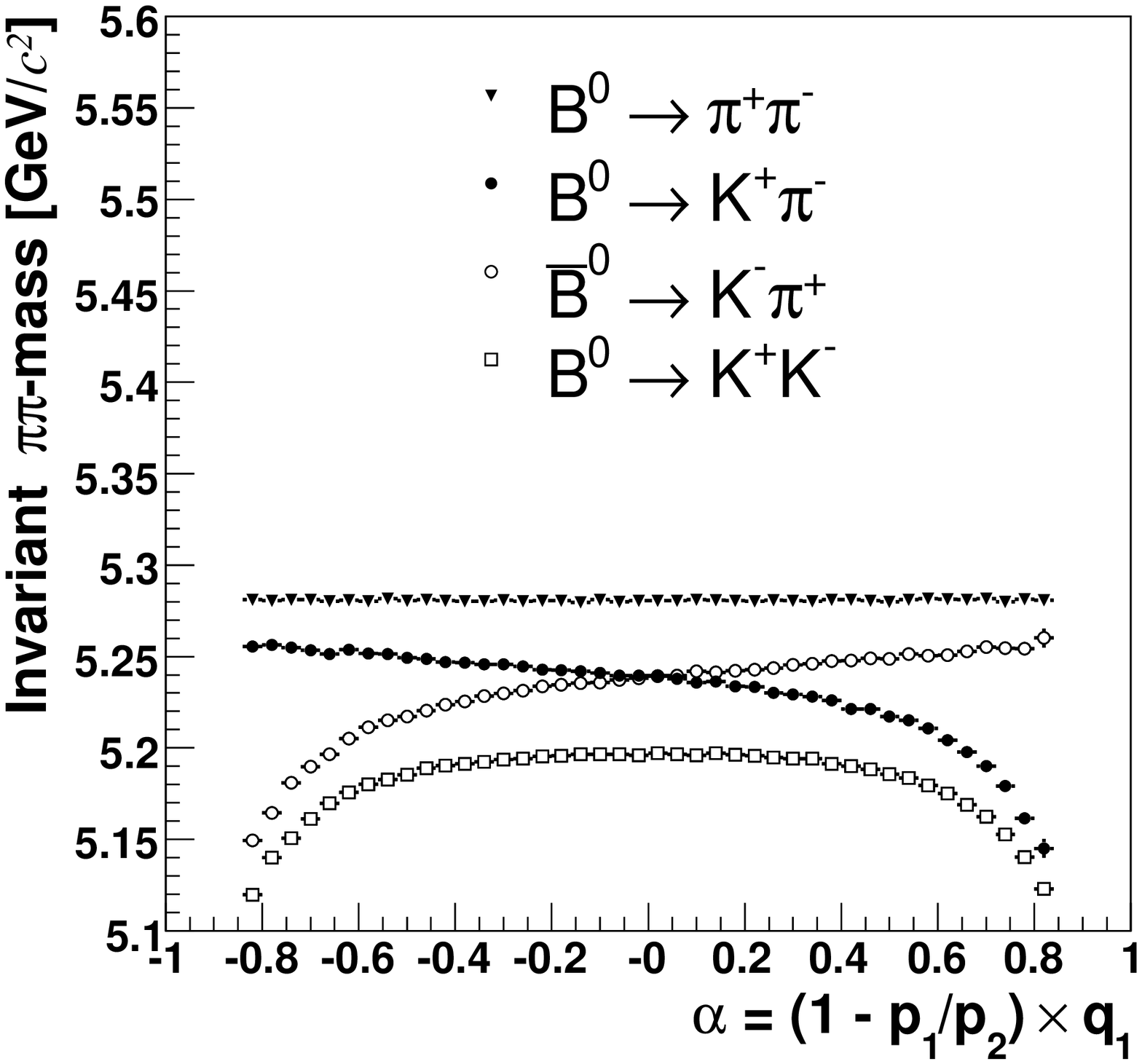}
\includegraphics[scale=0.22]{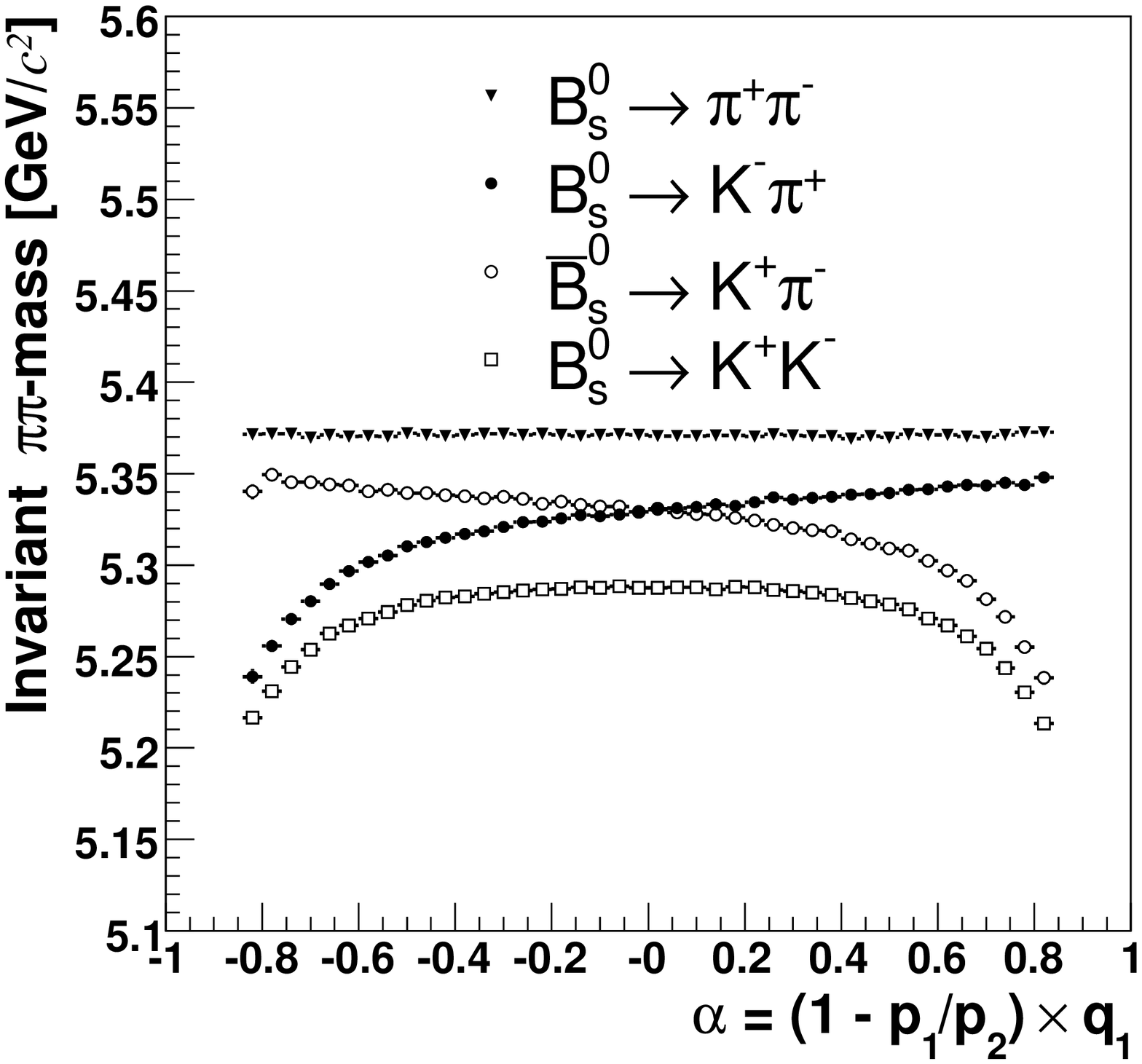}
\includegraphics[scale=0.22]{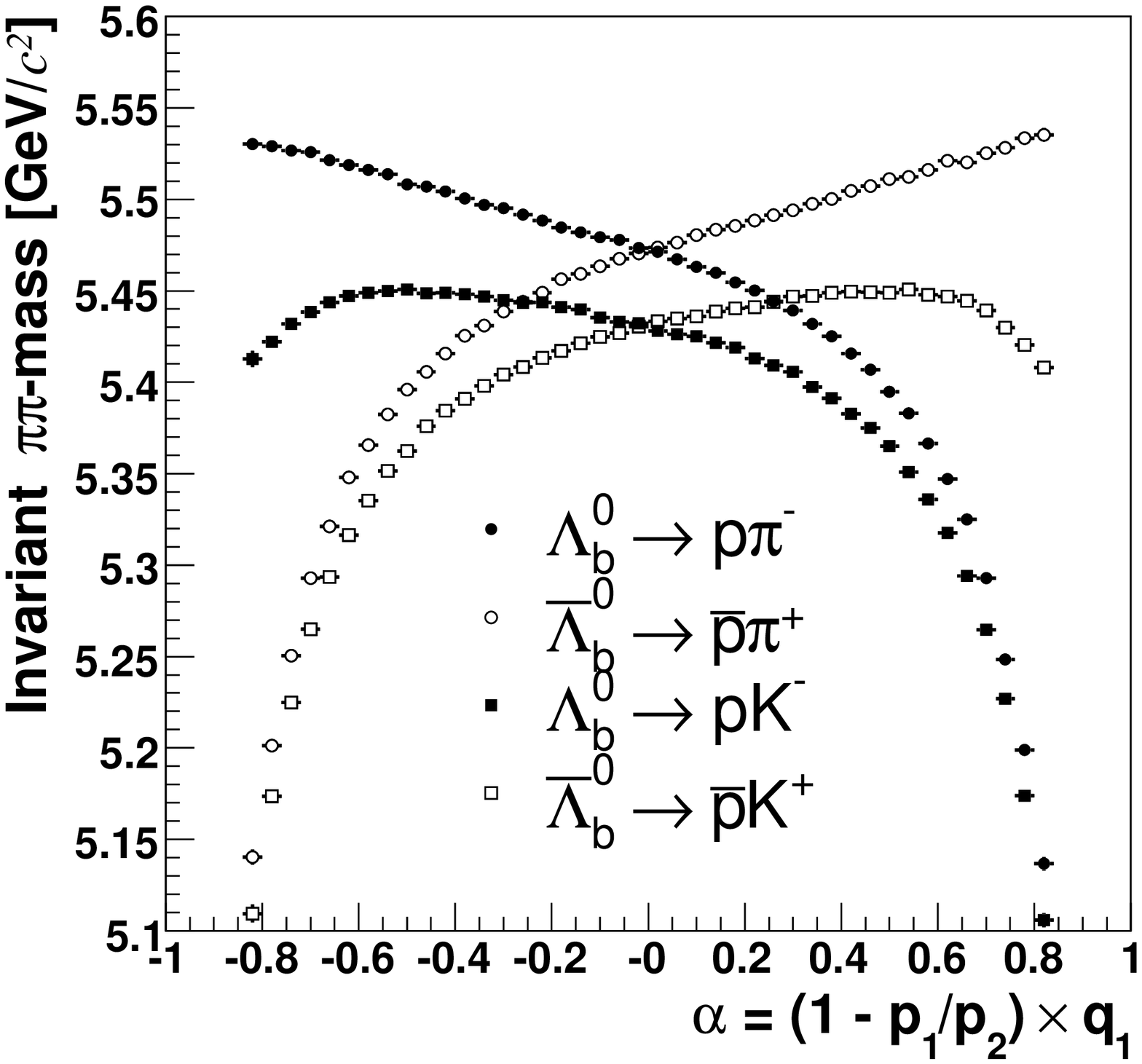}
\caption{Average \mpipi\ versus $\alpha$ for simulated
samples of \bd, \bs\ and \Lb\ decay modes.}
\label{fig:mpipi_vs_alpha1}
\end{center}
\end{figure}

\begin{figure}[htb]
\begin{center}
\includegraphics[scale=0.22]{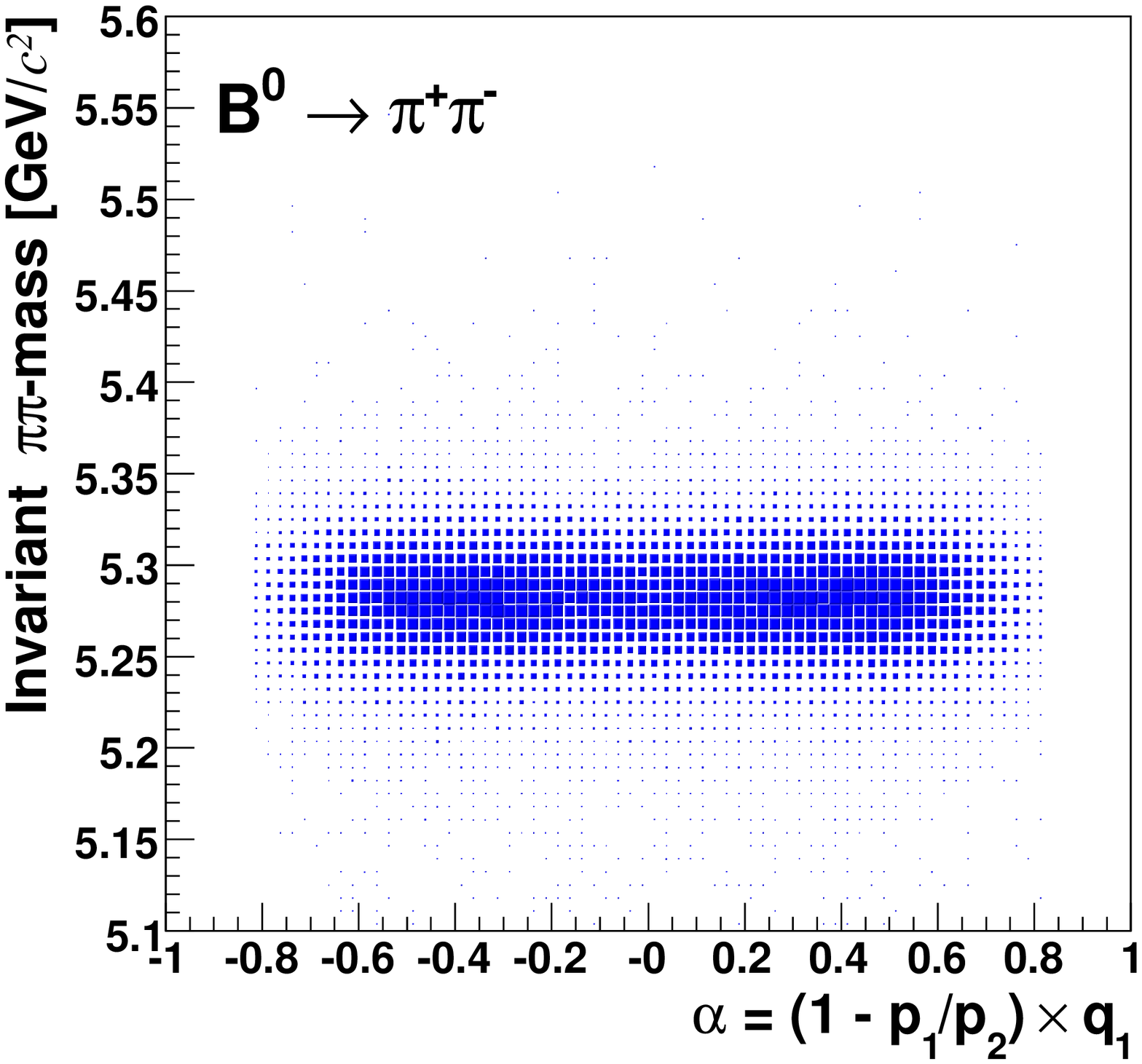}
\includegraphics[scale=0.22]{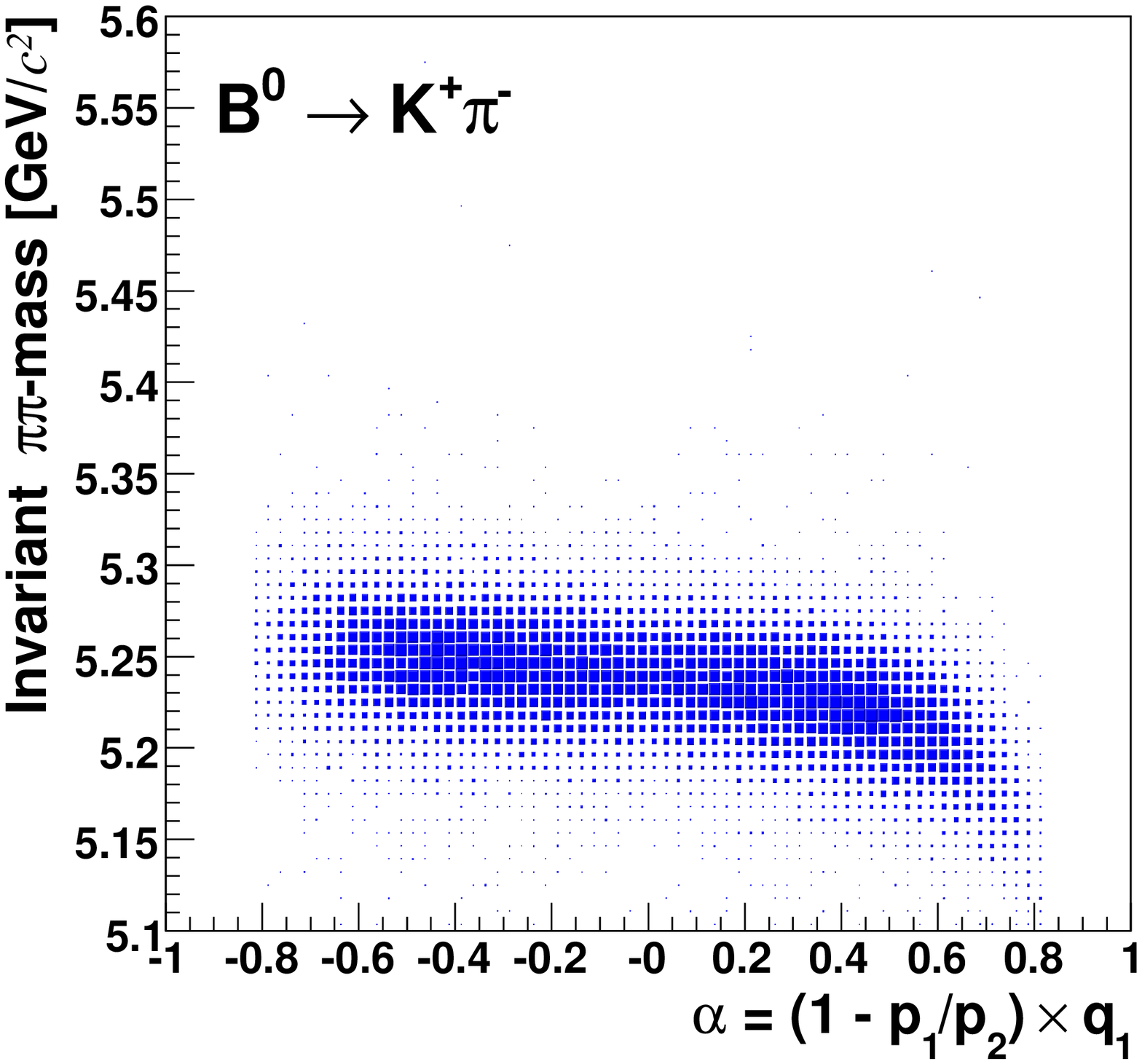}
\includegraphics[scale=0.22]{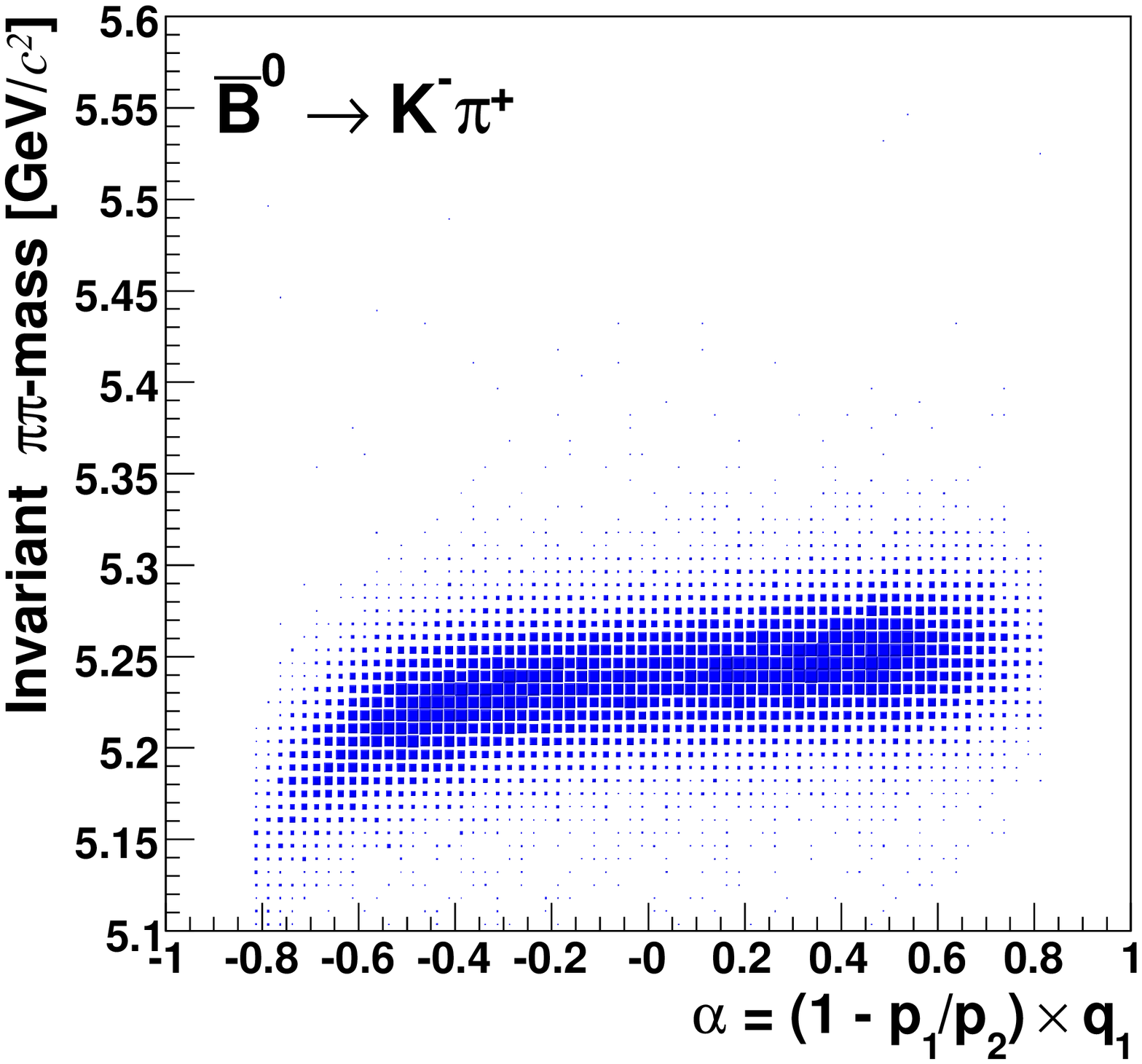}
\includegraphics[scale=0.22]{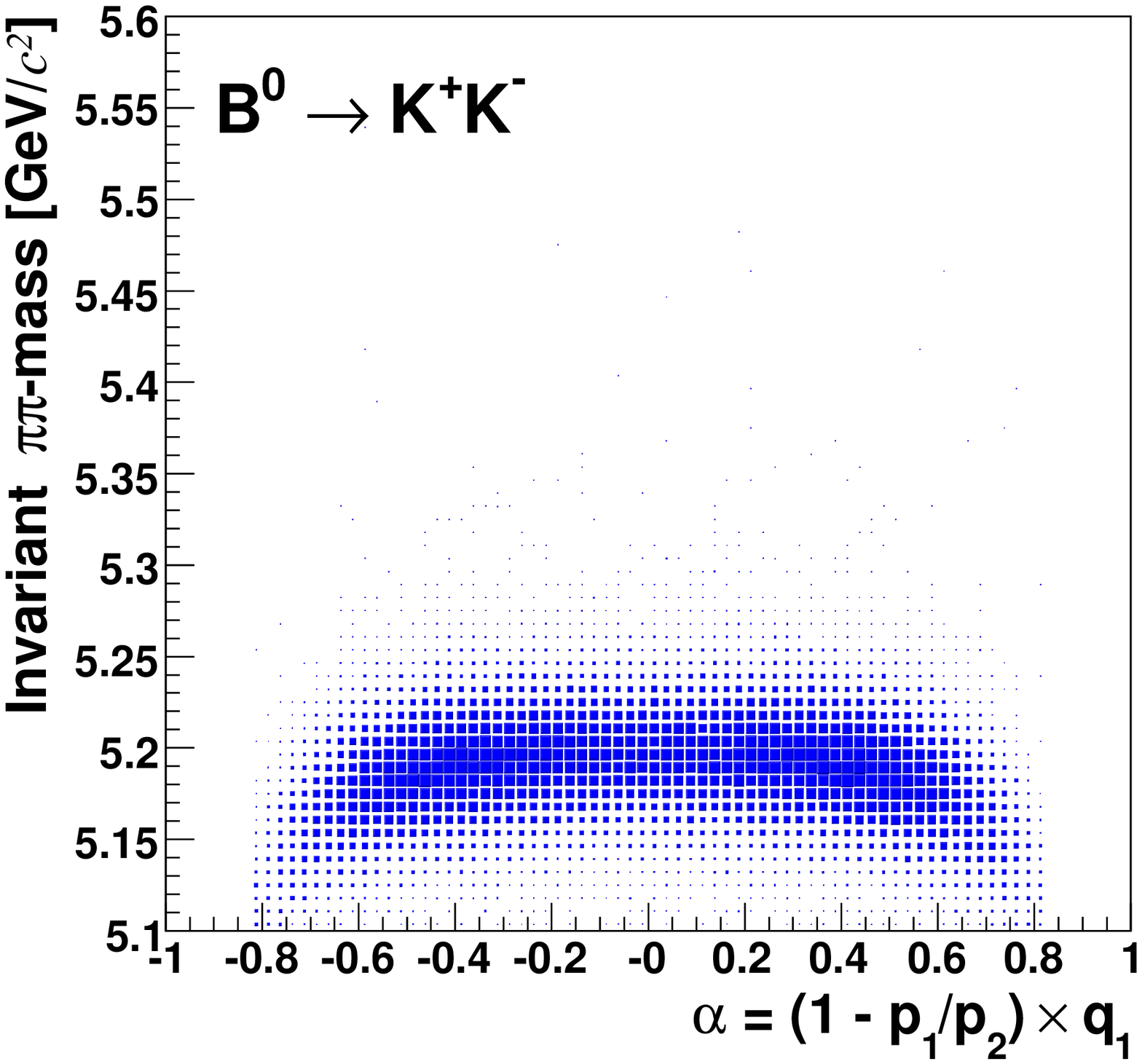}
\includegraphics[scale=0.22]{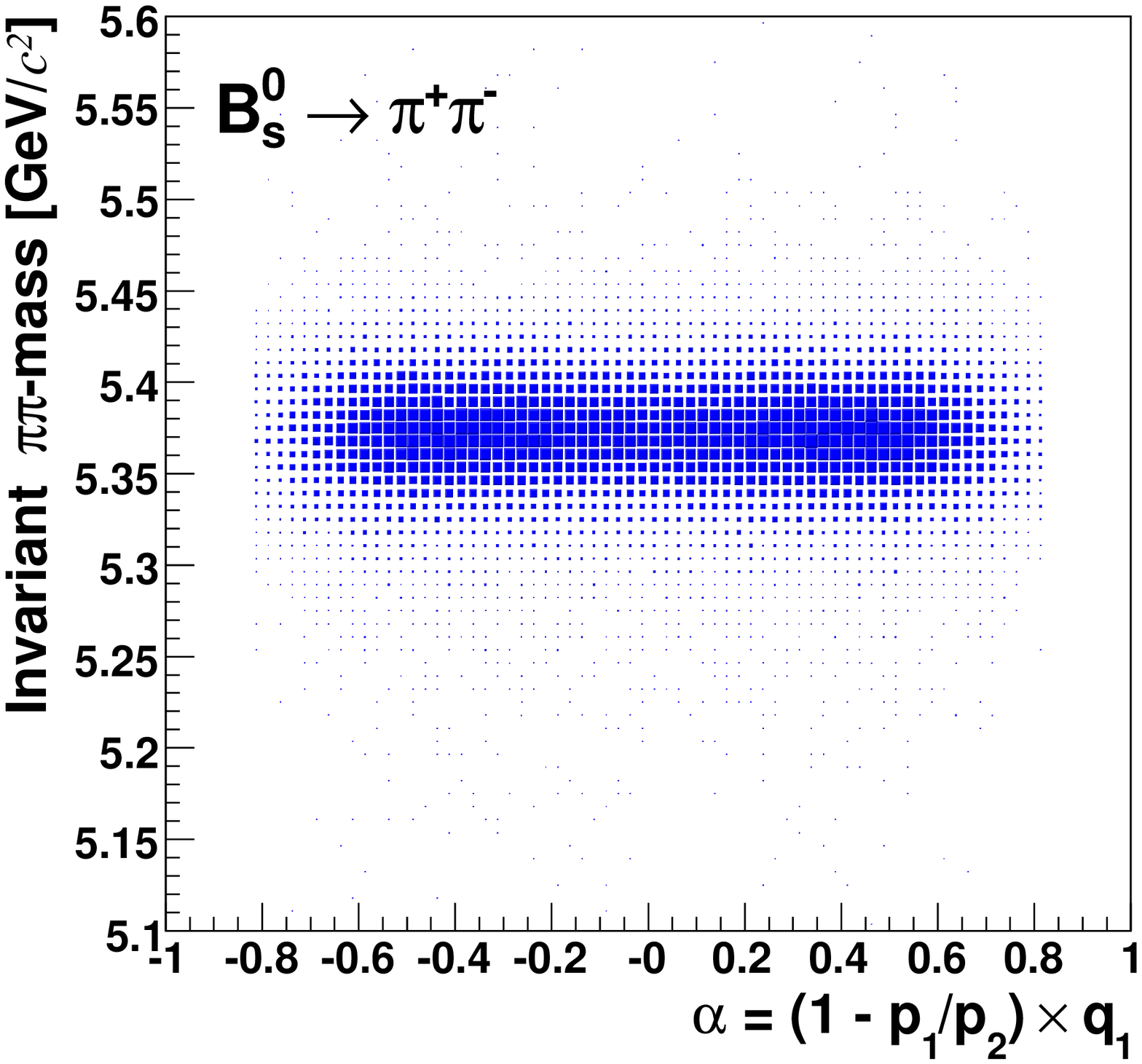}
\includegraphics[scale=0.22]{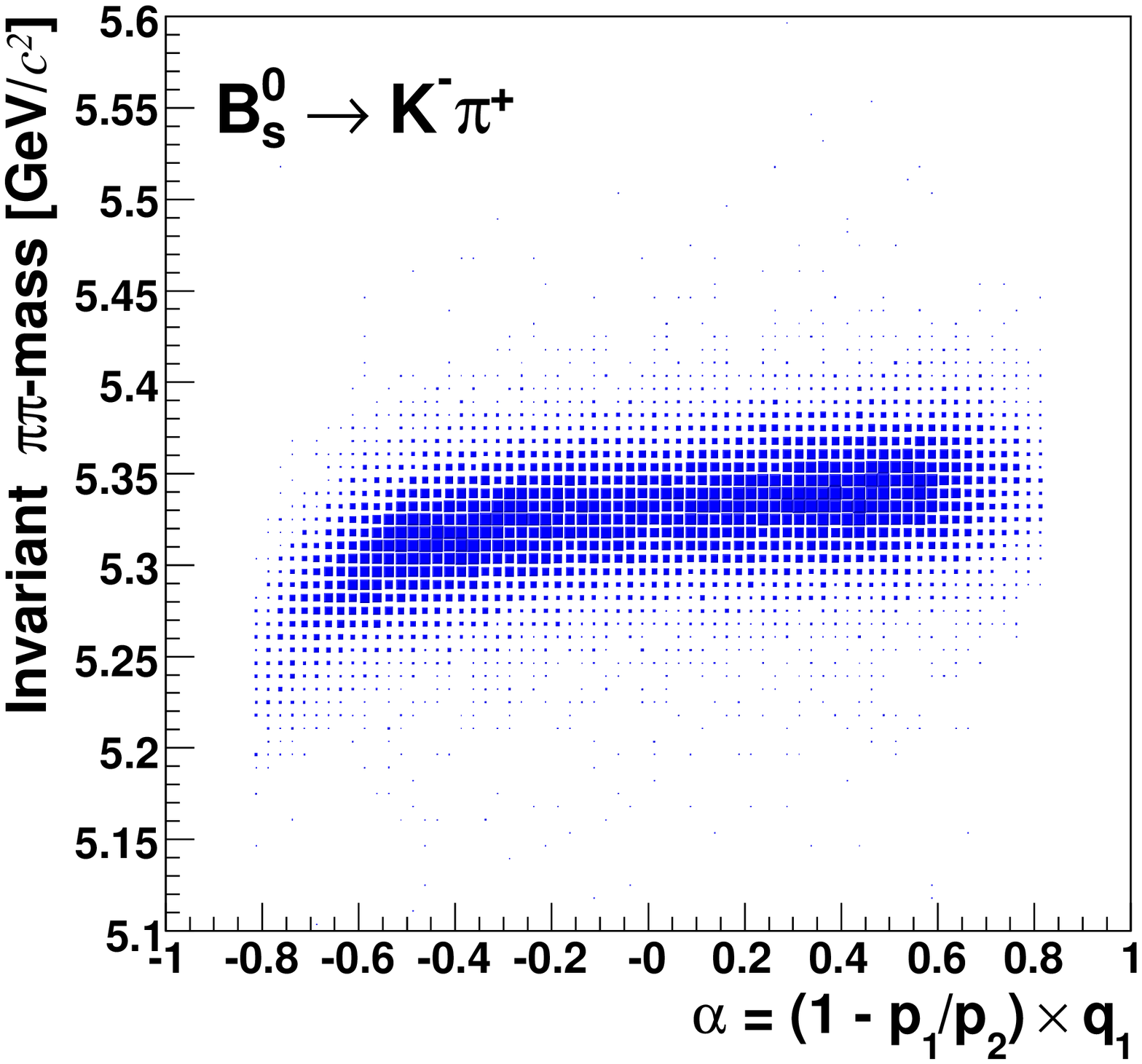}
\includegraphics[scale=0.22]{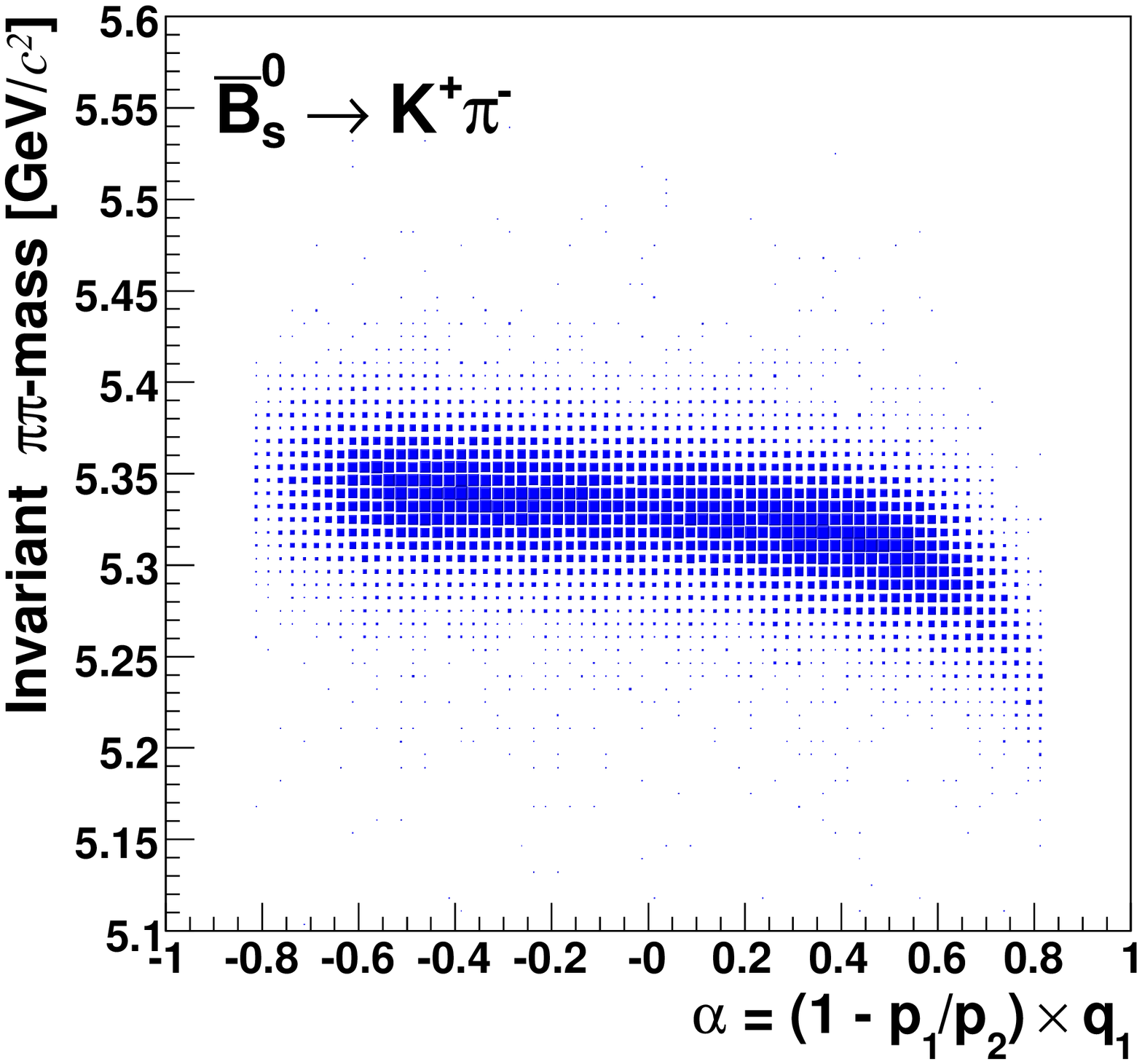}
\includegraphics[scale=0.22]{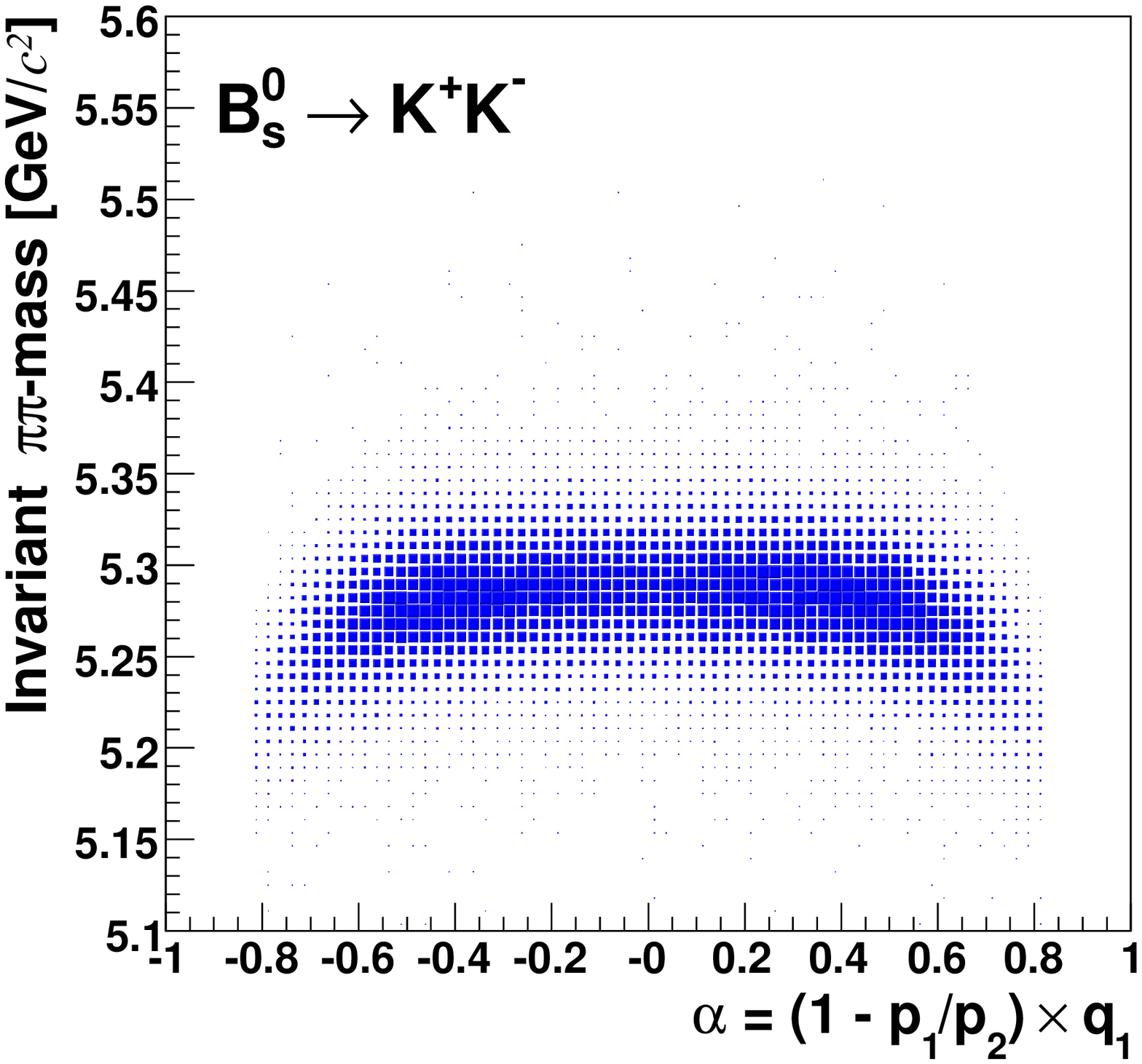}
\includegraphics[scale=0.22]{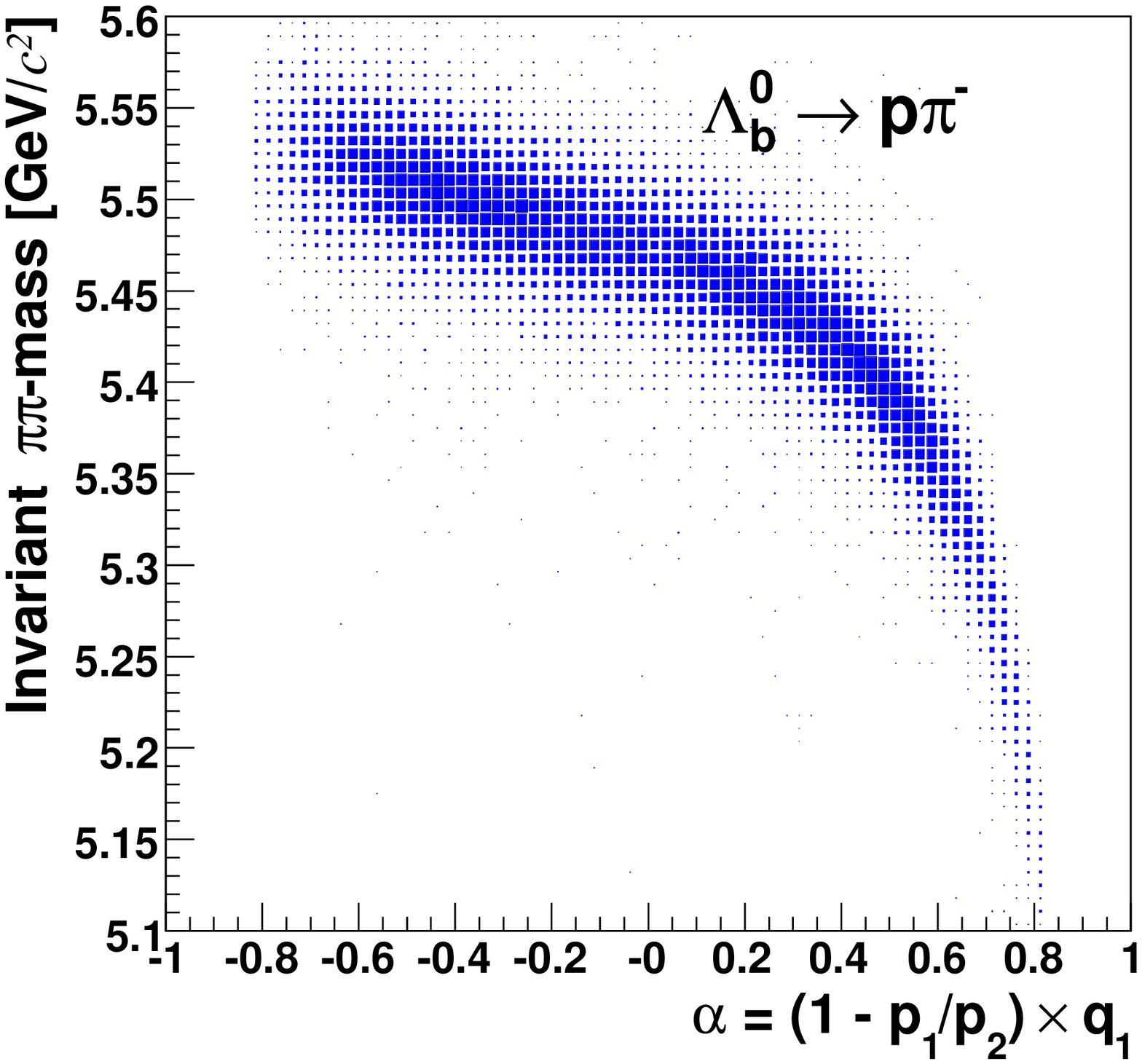}
\includegraphics[scale=0.22]{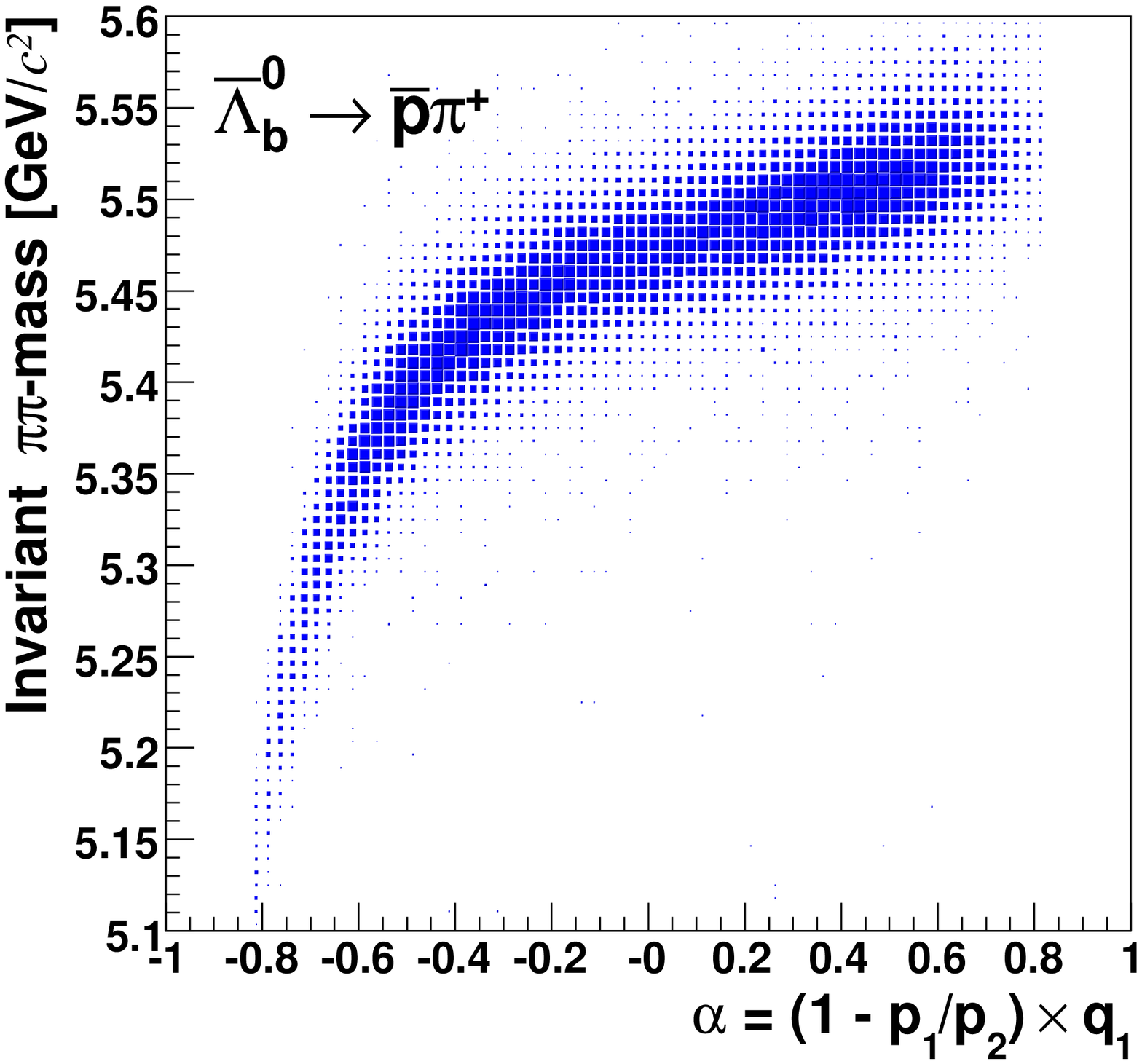}
\includegraphics[scale=0.22]{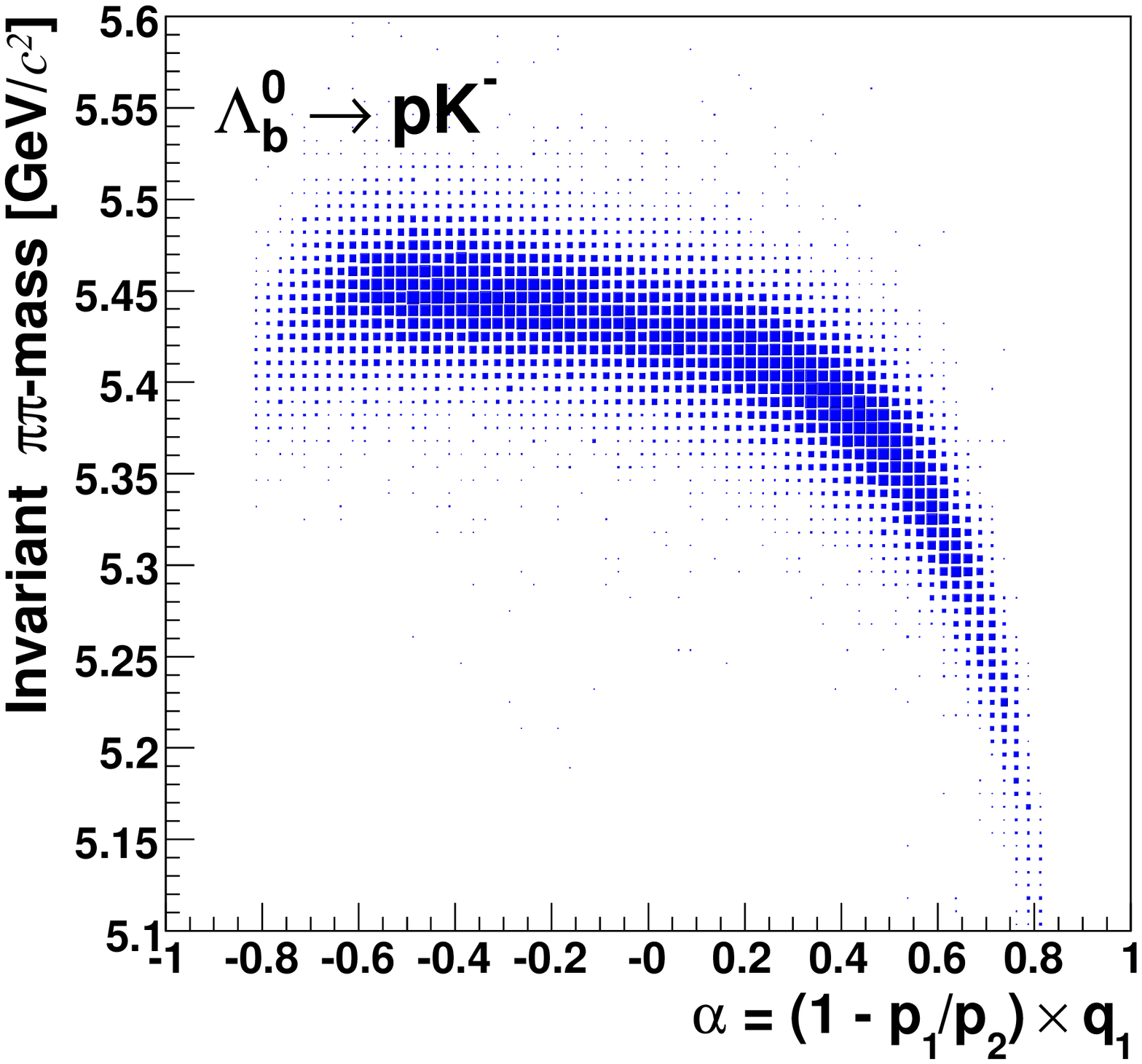}
\includegraphics[scale=0.22]{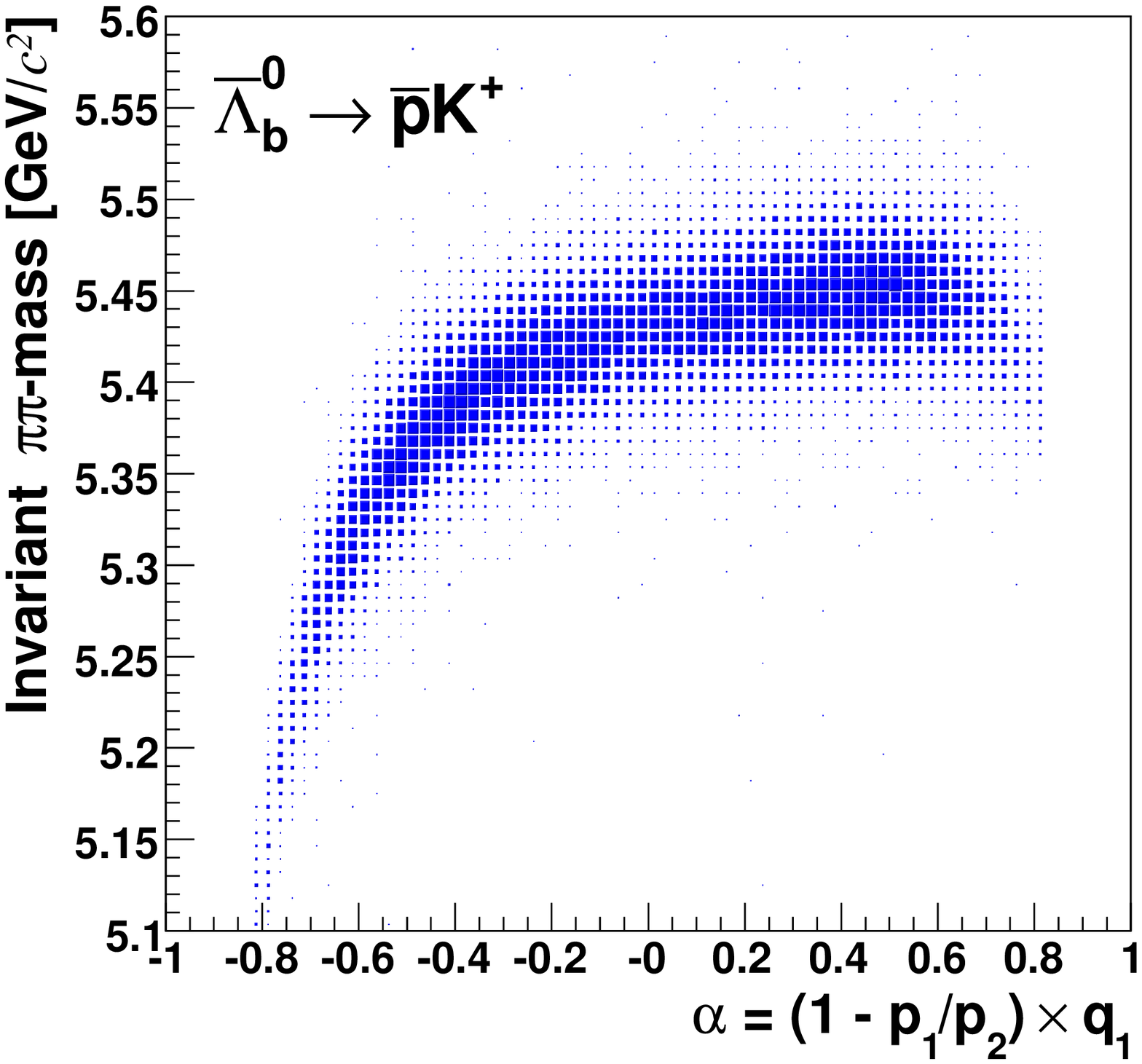}
\caption{\mpipi\ versus $\alpha$ for simulated samples of 
\Bhh\ and \Lbph\ decay modes.}
\label{fig:mpipi_vs_alpha2}
\end{center}
\end{figure}

The resolution in invariant mass and in particle identification is not 
sufficient for separating the individual decay modes on an event-by-event basis,
therefore we performed an unbinned maximum likelihood fit, combining kinematic and particle identification information
to statistically determine both the contribution of each mode,
and the relative contributions to the \CP\ asymmetries.
For the kinematic portion, we used three loosely correlated
observables to summarize the
information carried by all possible values of invariant mass of the
$b$-hadron candidate,
resulting from different mass assignments to the two outgoing
particles~\cite{punzibias}. They are: (a) the invariant $\pi\pi$-mass
\mpipi\ calculated with the charged pion mass assignment to both
particles; (b) the signed momentum imbalance
$\alpha = (1-p_1/p_2) q_{1}$, where $p_1$ ($p_2$) is the
lower (higher) of the particle momenta, and $q_1$ is the sign of the charge of the
particle of momentum $p_{1}$; (c) the scalar sum of the particle momenta $p_{tot}=p_1 + p_2$.
Using these three variables, the mass of any particular mode $m_{m_1m_2}$
($m_{K\pi}, m_{\pi K}, m_{KK}, m_{p\pi}, m_{\pi p}, m_{pK}, m_{Kp}$)
can be written  as:
\begin{eqnarray}\label{eq:Mpipi2}
m^{2}_{m_1m_2}  &= & m^{2}_{\pi\pi}  -  2 m_{\pi}^2 +(m_{1}^2+m_{2}^2)                                 \nonumber \\
            &  &                 -  2 \sqrt{p_{1}^2+m_{\pi}^2} \cdot \sqrt{p_{2}^2+m_{\pi}^2}      \nonumber \\
            &  &                 +  2\sqrt{p_{1}^2+m_{1}^2} \cdot \sqrt{p_{2}^2+m_{2}^2},
\end{eqnarray}
\vspace{-0.5cm}
\begin{equation}\label{eq:sostituzione}
p_1 = \frac{1-|\alpha|}{2-|\alpha|}p_{tot} ~,~p_2 = \frac{1}{2-|\alpha|}p_{tot},
\end{equation}
where $m_{1}$ ($m_{2}$) is the mass of the lower (higher) momentum
particle. For simplicity, Eq.~(\ref{eq:Mpipi2}) is written
as a function of $p_{1}$ and $p_{2}$,
but in the likelihood it was used as a function of $\alpha$ and $p_{tot}$. 
The simulated average values of \mpipi\ as a function of $\alpha$ for the twelve 
\bshh\ and \Lbph\ modes are shown in Fig.~\ref{fig:mpipi_vs_alpha1} and Fig.~\ref{fig:mpipi_vs_alpha2}.
Particle identification (PID) information  is
summarized by a single observable kaonness $\kappa_{1(2)}$ for track $1(2)$, 
defined as 
$$ \kappa_{1(2)} = \frac{dE/dx_{1(2)} - dE/dx_{1(2)}(\pi)}{dE/dx_{1(2)}(K) - dE/dx_{1(2)}(\pi)},$$
where $dE/dx_{1(2)}(\pi)$ and $dE/dx_{1(2)}(K)$ are the expected $dE/dx_{1(2)}$ depositions for those particle assignments.
With the chosen observables, the likelihood
 contribution of the $i^{\rm th}$ event is written as:
\begin{equation}\label{eq:likelihood}
    \mathcal{L}_i  =  (1-f_b)\sum_{j} f_j \mathcal{L}^{\mathrm{kin}}_j  \mathcal{L}^{\mathrm{PID}}_j
                   +  f_b \left( f_{\rm{A}} \mathcal{L}^{\mathrm{kin}}_{\mathrm{A}}
    \mathcal{L}^{\mathrm{PID}}_{\mathrm{A}}+
   (1-f_{\rm{A}}) \mathcal{L}^{\mathrm{kin}}_{\mathrm{C}}
    \mathcal{L}^{\mathrm{PID}}_{\mathrm{C}}
	\right)
\end{equation}
where:
\begin{equation}
    \label{eq:signal}\mathcal{L}_j^{\mathrm{kin}}=
     R_{j}(\mpipi|\alpha,p_{tot})
     P_{j}(\alpha,p_{\rm{tot}}),
\end{equation}
 \begin{equation}\label{eq:bck_A}
\mathcal{L}^{\mathrm{kin}}_{\mathrm{A}} =
           {\rm A}(\mpipi;c_{2},m_{0}) P_{\mathrm{A}}(\alpha,p_{\rm{tot}}),
\end{equation}
\begin{equation}
\label{eq:bck_E}\mathcal{L}^{\mathrm{kin}}_{\mathrm{C}} =
           e^{c_{1} \mpipi} P_{\mathrm{C}}(\alpha,p_{\rm{tot}}),
\end{equation}
\begin{equation}
\label{eq:PID_sig} \mathcal{L}^{\mathrm{PID}}_{j} =
   	F_{j}(\kappa_{1}, \kappa_{2}|\alpha,p_{\rm{tot}}),
\end{equation}
\begin{equation}
\label{eq:PID_bg} \mathcal{L}^{\mathrm{PID}}_{\mathrm{A(C)}} =
        \sum_{l,m=e,\pi,K,p} w^{\mathrm{A(C)}}_{l} w^{\mathrm{A(C)}}_{m} F_{lm}(\kappa_{1}, \kappa_{2}|\alpha,p_{\rm{tot}}).
\end{equation}
The various terms of the likelihood functions are described below.

The index $j$ runs over the twelve distinguishable \Bhh\ and \Lbph\ modes, and
$f_j$ are their fractions to be determined by the fit, together with the total background fraction $f_b$.
The background is composed of two different kinds: combinatorial background and 
partially-reconstructed heavy flavor decays. The combinatorial background is composed of random pairs 
of charged particle, displaced from the beam-line, accidentally satisfying the selection requirements,
while the  latter, referred as ``physics'' background, is composed of multi-body $b$-hadron decays
(i.e.~$\bn \to \rho\pi/\rho K$) in which only two tracks are reconstructed.   
The indices $\mathrm{A(C)}$ label the physics (combinatorial) background quantities.
The fraction of the physics background is given by $f_{A}$ and it is a free parameter in the fit.

Each likelihood term, both for signals and backgrounds, is factorized into three different contributions:
a) the conditional probability distribution of the invariant mass \mpipi\ given
$\alpha$ and $p_{tot}$ (for the background \mpipi\  is assumed to be independent of momentum), 
b) the joint conditional probability of PID variables 
$\mathcal{\kappa}_1$, $\mathcal{\kappa}_2$ given $\alpha$, $p_{tot}$ for a determined particles hypothesis, $j$ in the case of 
signals ($F_{j}$) and $l,m$ in the case of background ($F_{l,m}$), and 
c) the joint probability distribution of momentum variables $\alpha$ and $p_{tot}$ ($P_{j\mathrm{(A,C)}}$).

\begin{figure}[tb]
\begin{overpic}[scale=0.33]{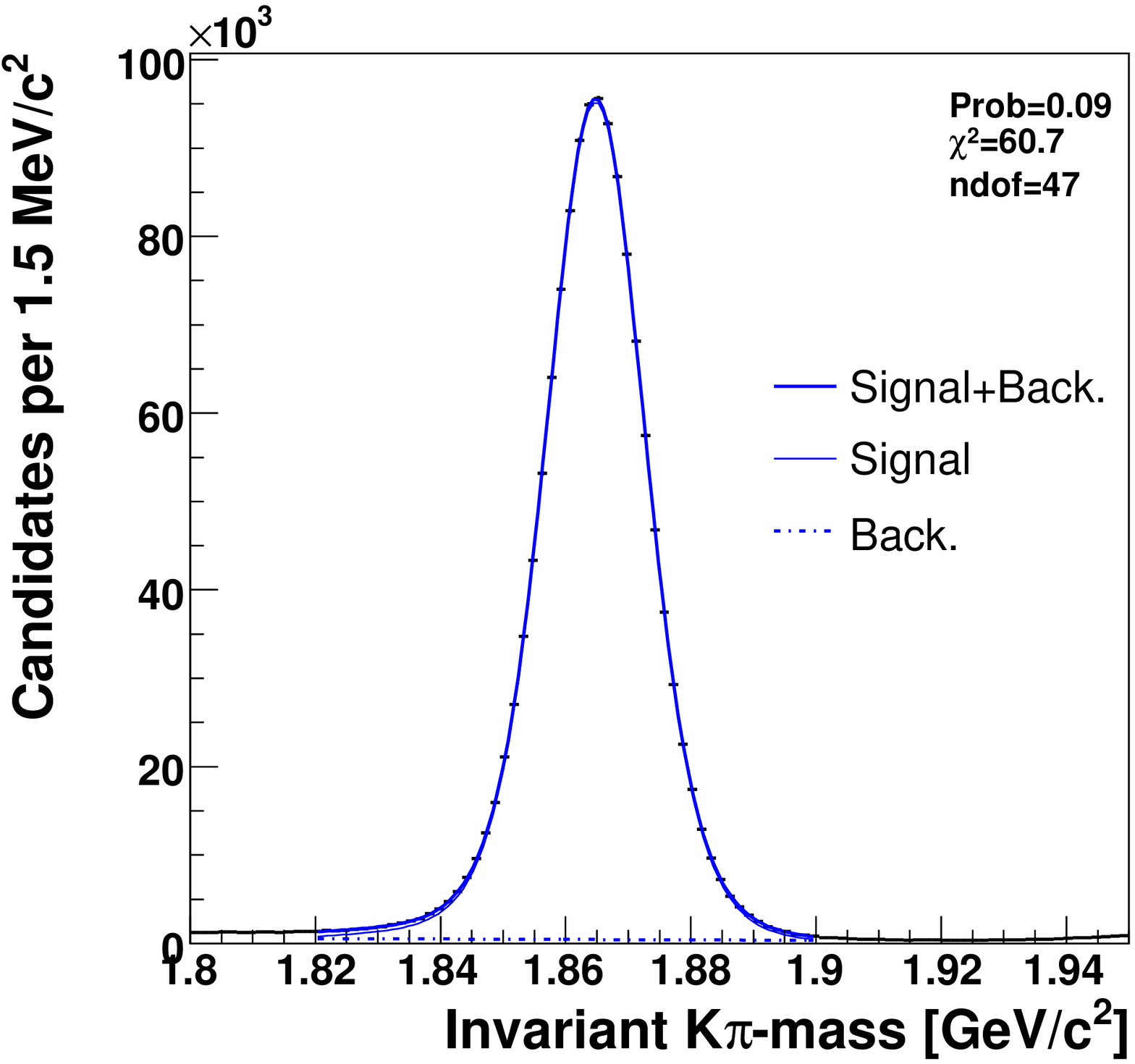}
\put(33,150){(a)}
\end{overpic}   
\begin{overpic}[scale=0.33]{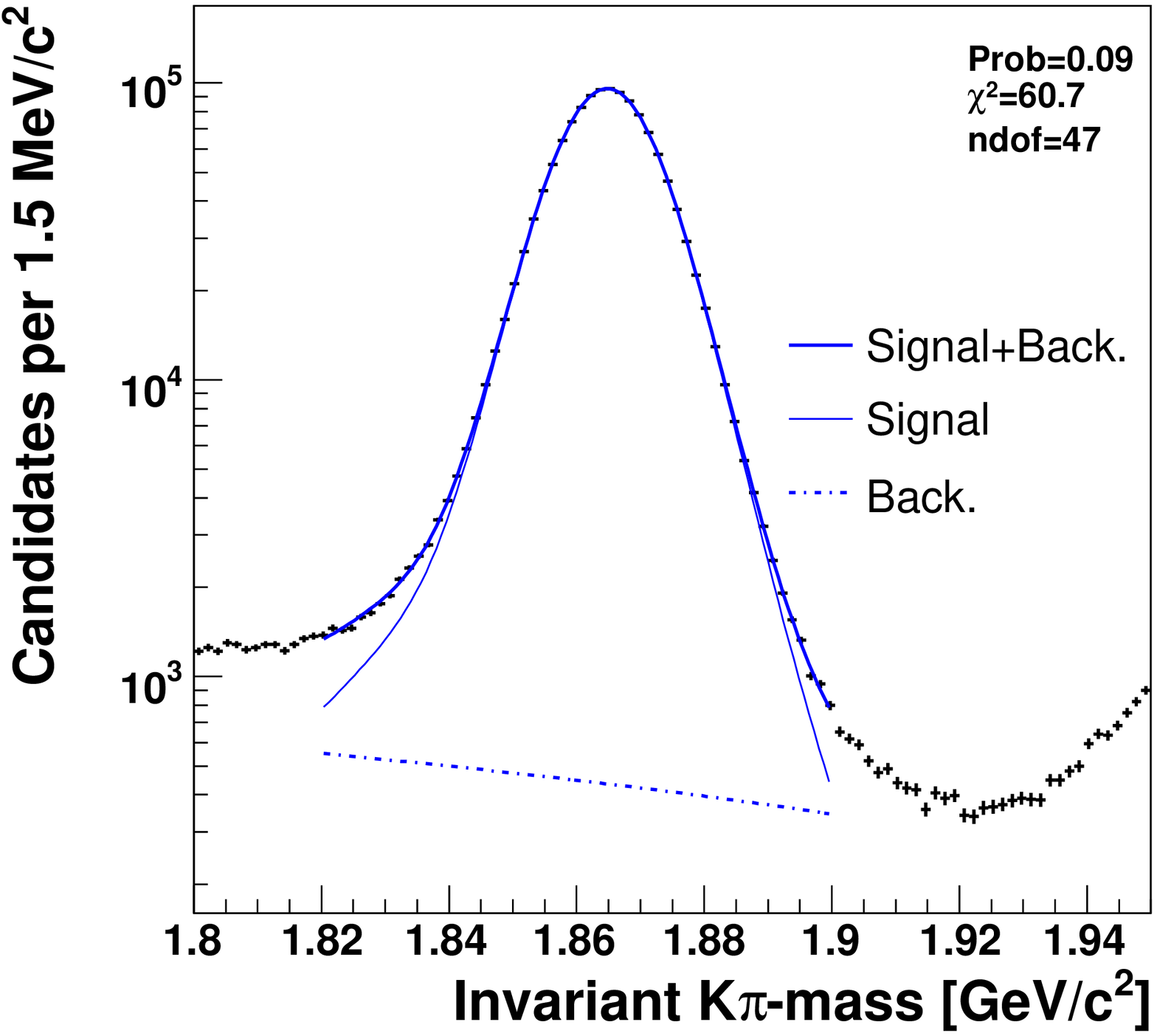}
\put(33,150){(b)}
\end{overpic}   
\caption{Invariant-$K\pi$ mass distribution for tagged \Dkpi\ decays 
from $\ensuremath{D^{*+} \to D^{0} \pi^{+} \to [K^{-}\pi^{+}]\pi^{+}}$. 
A verification of the mass line shape is superimposed,
by performing a 1-D binned fit where the signal mass line shape is completely fixed from the model (see text).
Linear scale (a), logarithmic scale (b).
}
\label{fig:mass_resol}
\end{figure}

\begin{figure}[tb]
\begin{overpic}[scale=0.33]{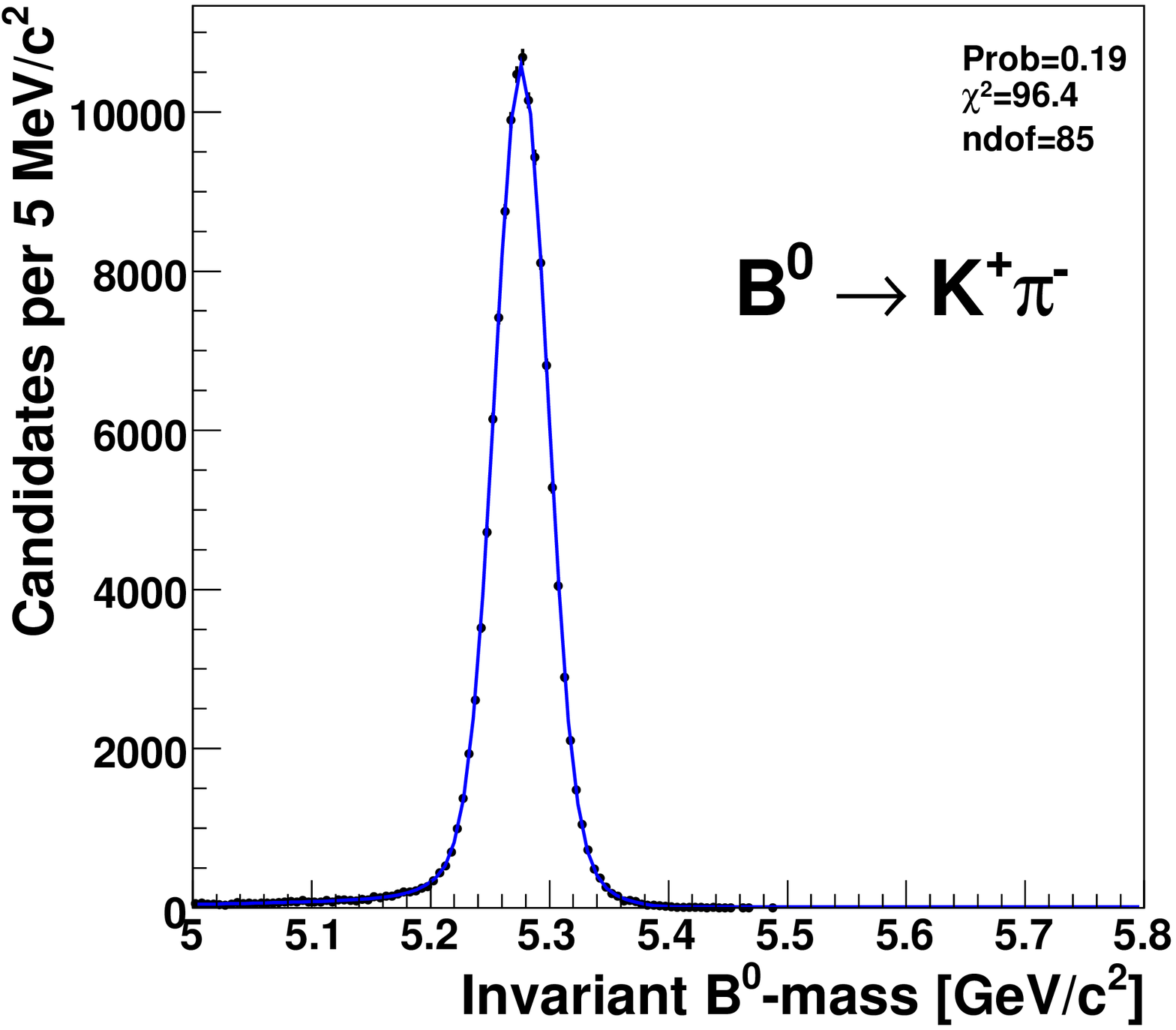}
\put(33,150){(a)}
\end{overpic}   
\begin{overpic}[scale=0.33]{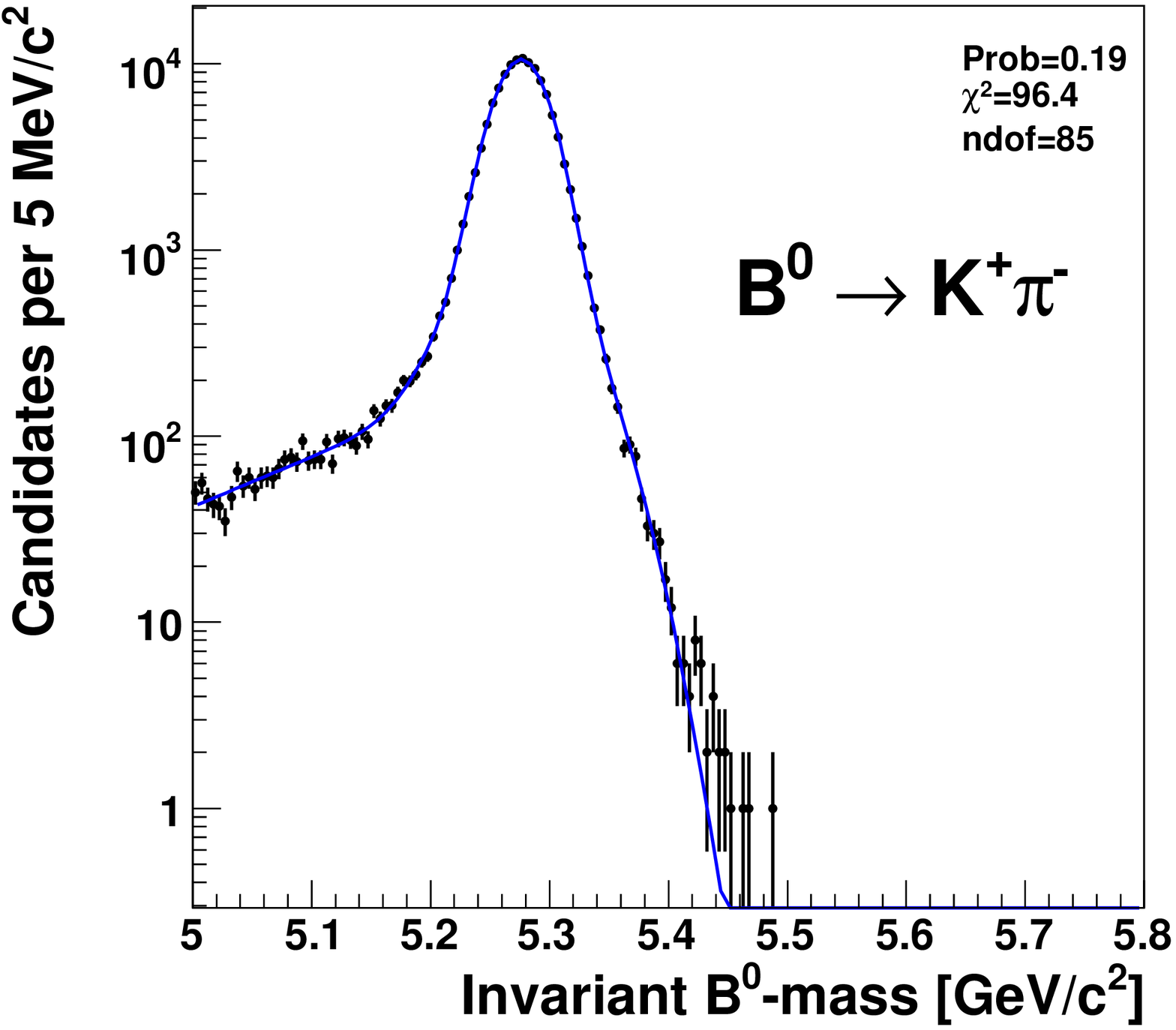}
\put(33,150){(b)}
\end{overpic}   
\caption{Invariant-$K\pi$ mass distribution of simulated \BdKpi\ decays. 
The mass template is superimposed. Linear scale (a), logarithmic scale (b).
Similar templates for all \Bhh\ and \Lbph\ decay modes.
}
\label{fig:mass_resol_templates}
\end{figure}

If $\mathcal{R}^{j}(m_{j})$ is the mass resolution function of each mode $j^{\rm th}$ when the correct mass is assigned to both tracks,
we can use Eq.~(\ref{eq:Mpipi2}) to change variable $m_{j} \to \mpipi$ and to write the density probability function 
for each $j^{\rm th}$ decays mode as function of \mpipi\ given $\alpha$ and $p_{tot}$. In fact:
\begin{equation}
\mathcal{R}_{j}(m_{j})=\mathcal{R}_{j}(m_{j}(\mpipi)) \cdot \frac{d\mpipi}{dm_{j}}=R_{j}(\mpipi|\alpha,p_{tot}).
\end{equation}
The functional form of the mass resolution function $\mathcal{R}_{j}(m_{j})$  
was parameterized using the detailed detector simulation. To take into account non-Gaussian tails 
due to the emission of photons in the final state, we included  soft photon emission in the simulation, 
using recent QED calculations \cite{Cirigliano-Isidori}.
The quality of the mass resolution model was verified by comparison data and simulation 
with about $1.5 \times 10^{6}$ tagged \DKpi\ decays reconstructed using the chain $D^{*+}\to D^0\pi^+\to [K^-\pi^+]\pi^+$
(see Fig.~\ref{fig:mass_resol} and  Fig.~\ref{fig:mass_resol_templates}). 
The mass line-shape of the \DKpi\ was fitted by fixing the signal 
shape from the model, and allowing to vary only the background function.
Good agreement was obtained between data and simulation.
In Eq.~(\ref{eq:signal}),  the nominal \Bd, \Bs\ and \Lb\ masses measured by CDF~\cite{CDFmasspaper} were used
to reduce the systematic uncertainties related to the knowledge of the global mass scale.

The mass distribution of the physics background is parameterized with  
an ``Argus function'', defined by the notation
${\rm A}(\mpipi;c_{2},m_{0})$~\cite{argus}, convoluted with a Gaussian distribution
centered at zero with a width, in this case, equal to the mass resolution,
while the combinatorial background with an exponential function. 
The background mass distribution was determined in
the fit by varying the parameters $c_1$, $c_2$ and $m_{0}$ in Eq.~(\ref{eq:bck_A},\ref{eq:bck_E}).
The function $P_{j\rm{(A,C)}}(\alpha,p_{tot})$ 
was parameterized by a product of
polynomial and exponential functions fitted to Monte Carlo samples
produced by a detailed detector simulation for each mode $j$, instead for the background 
 terms was obtained from the mass sidebands of data~\cite{my_thesis}.
\begin{figure}[tb]
\begin{overpic}[scale=0.35]{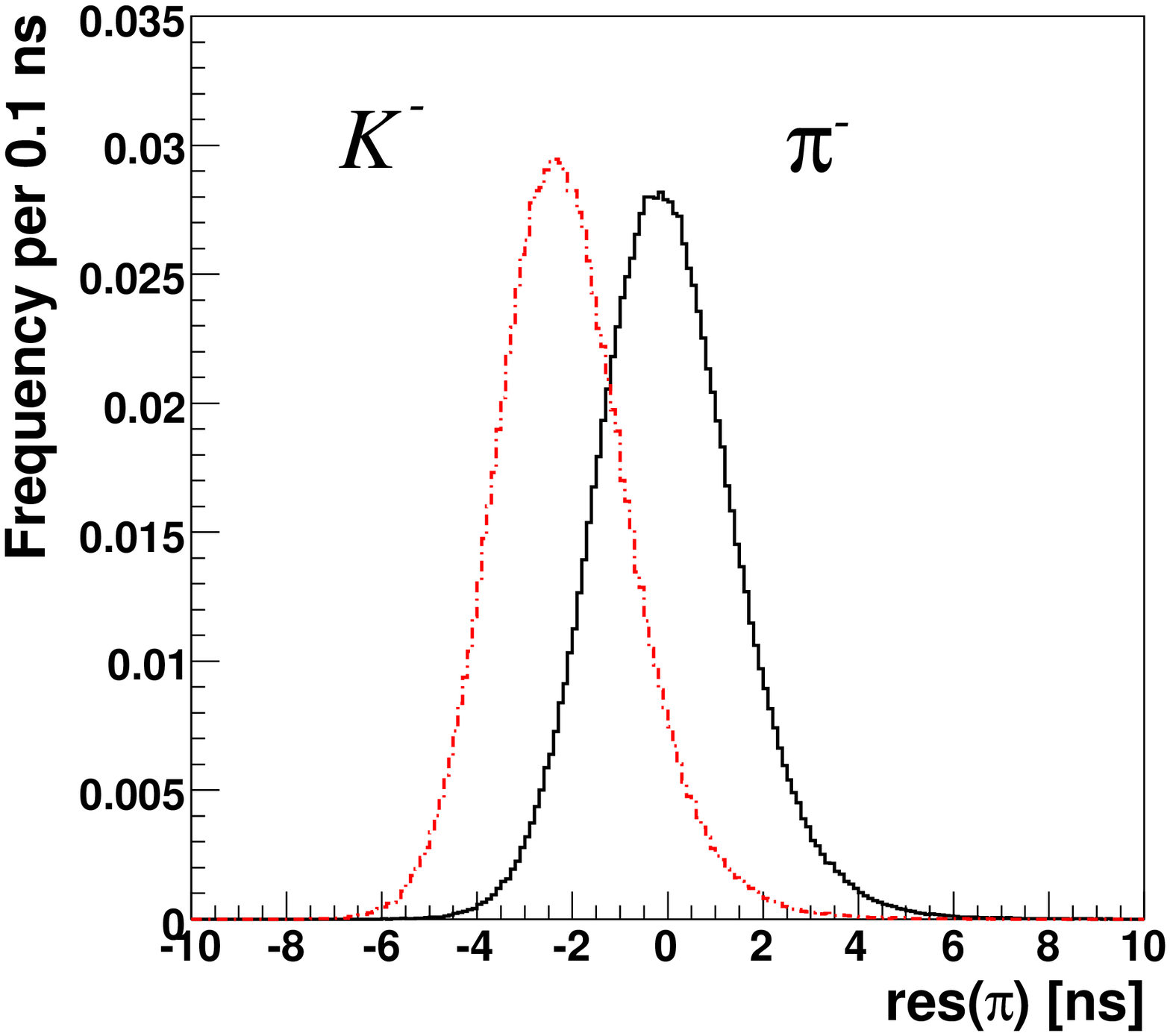}
\put(35,160){(a)}
\end{overpic}   
\begin{overpic}[scale=0.35]{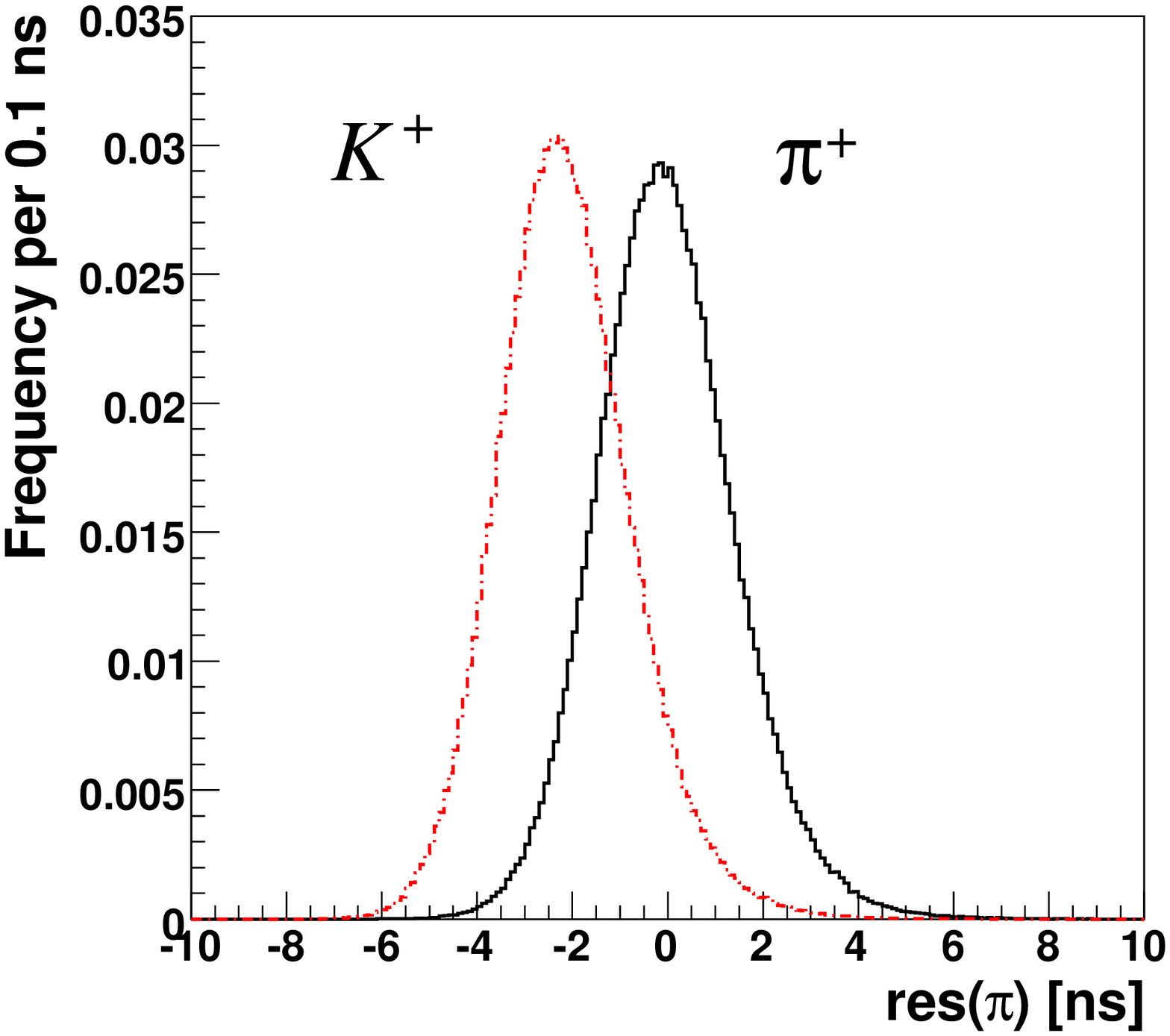}
\put(35,160){(b)}
\end{overpic}   
\caption{Distribution of \dedx\ 
around the average pion response for negatively- (a)
 and positively- (b) charged particles. Pions (continuous line) and kaons (dashed line) from \DKpi\
decays.}
\label{fig:dstar2}
\end{figure}
\begin{figure}[tb]
\includegraphics[scale=0.35]{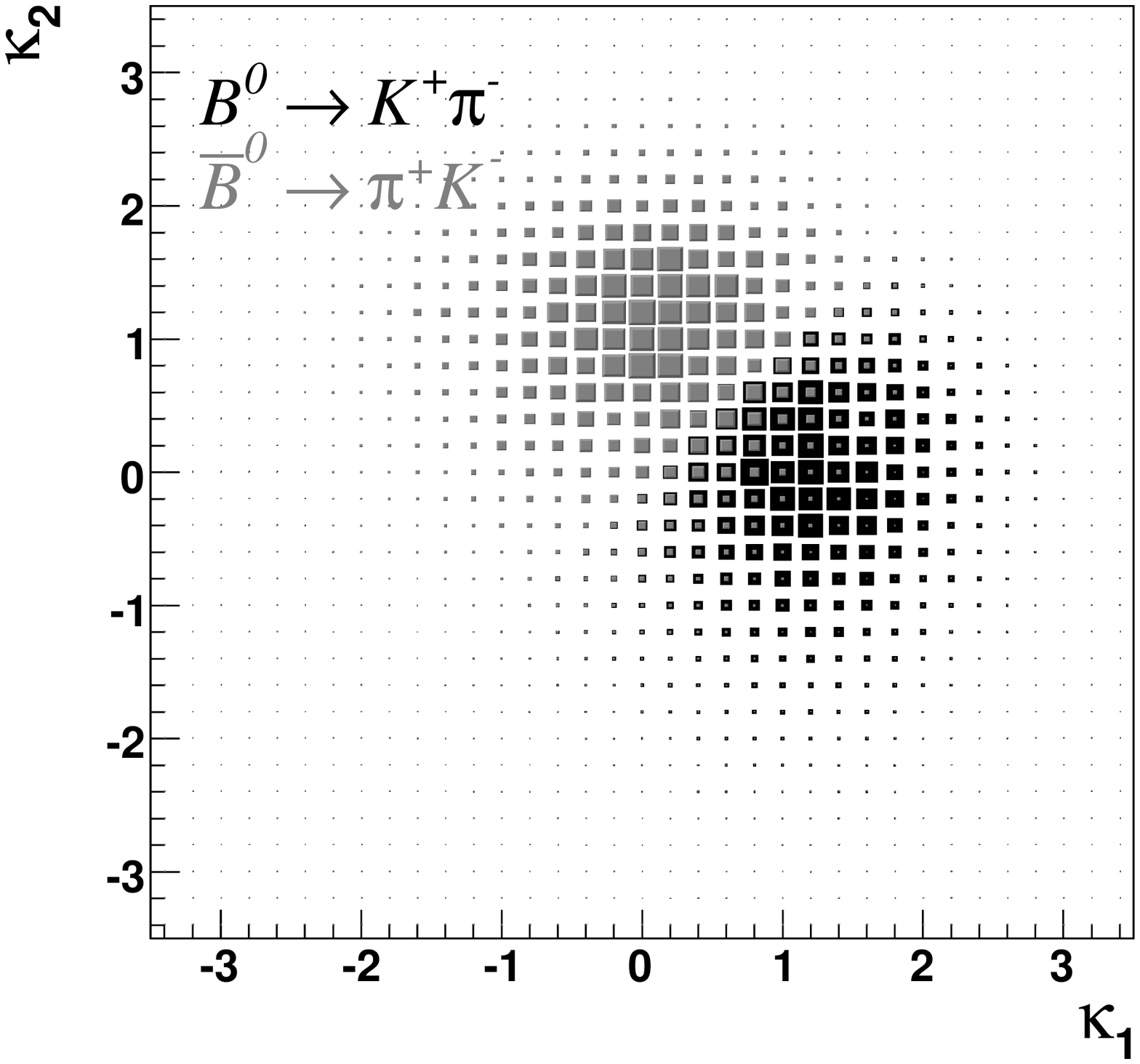}
\includegraphics[scale=0.35]{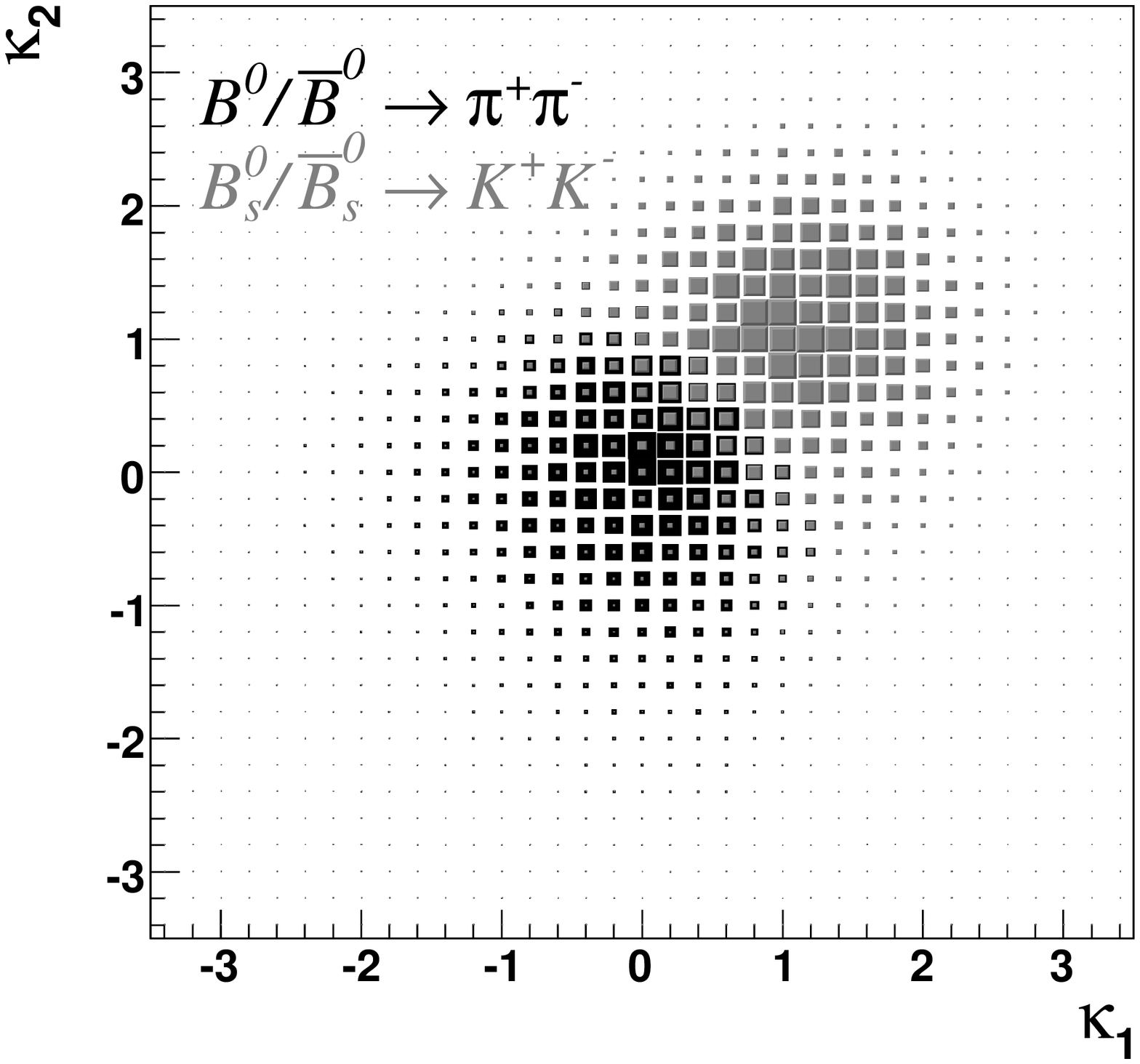}
\caption{Probability density function $\int F_{j}(\kappa_{1},\kappa_{2}|\alpha,p_{tot}) P_{j}(\alpha,p_{tot}) d\alpha d p_{tot}$ for 
the some signal decays in the space $\alpha>0$. To obtain the distributions for $\alpha<0$
it is sufficient to invert $\kappa_{1} \leftrightarrow \kappa_{2}$}
\label{fig:id_2d_sov}
\end{figure}

The same data sample of $1.5 \times 10^{6}$ $D^{*+}\to D^0\pi^+\to [K^-\pi^+]\pi^+$ decays
used to test mass resolution model, where the $D^0$ decay
products are identified by the charge of the $D^{*+}$ pion,
was used to calibrate the \dedx\ response over the tracking volume
and over time, and to determine the $F_{j(l,m)}(\mathcal{\kappa}_1,\mathcal{\kappa}_2|\alpha,p_{tot})$ functions
in Eq.~(\ref{eq:PID_sig},\ref{eq:PID_bg}). 
In a $>95\%$ pure $D^0$ sample, we obtained approximately $1.5\sigma$ separation between kaons and pions
for particles with momentum larger than 2~\pgev\ (see Fig.~\ref{fig:dstar2}), 
corresponding to an uncertainty on the measured fraction of each class 
of particles that is just 1.7 times worse than the uncertainty  attainable with ideal separation between 
two classes of events completely disentangled.
The effective separation among final states consisting in particle pairs,
like in our case (between $\pi^+\pi^-$  and $K^+K^-$, 
between $\pi^+K^-$  and $K^+\pi^-$) corresponds  to $1.5\sigma \cdot\sqrt{2} \simeq 2.1\sigma$, as shown 
in Fig.~\ref{fig:id_2d_sov}.
This achievement is particularly crucial in separating those signal decay modes in which the kinematics 
does not sufficiently  help. 
For example the kinematic separation power between the \Bdpipi\ and \BsKK\ modes is almost null, 
as shown in Fig.~\ref{fig:mpipi_vs_alpha1} and Fig.~\ref{fig:mpipi_vs_alpha2},
while the \dedx\ power separation is maximum, 
about $2.1\sigma$, as shown in Fig.~\ref{fig:id_2d_sov}. 

The \dedx\ response of protons was determined from a sample of 124,000 $\Lambda\to p \pi^{-}$ decays, 
where the kinematics and the momentum threshold of the trigger allow unambiguous 
identification of the decay products~\cite{my_thesis}.

The PID background term in Eq.~(\ref{eq:PID_bg}) 
is similar to the signal terms, but allows for independent pion, kaon, proton, and electron components, 
which are free to vary independently for physics (combinatorial) background. 
In Eq.~(\ref{eq:PID_bg}) the indices $l$ and $m$ run over the four possible particles 
$e,~\pi,~K,~p$ and the fractions of different kind of particles $w^{\rm{A(E)}}_{l}$,$w^{\rm{A(E)}}_{m}$
are free parameters in the fit. Muons are indistinguishable from pions with the
available \dedx\ resolution.

\begin{figure}[htb]
\begin{overpic}[scale=0.22]{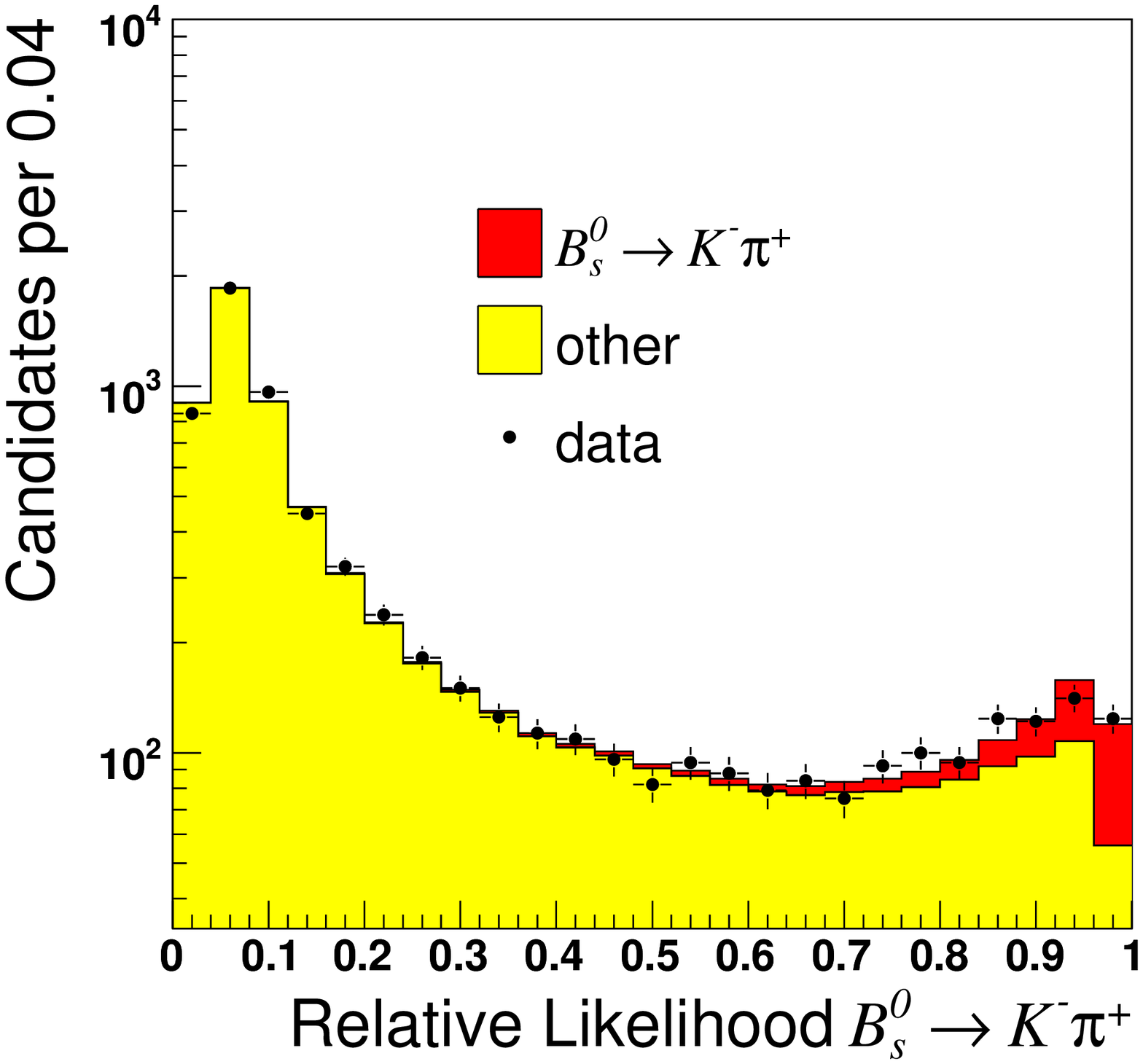}
\put(25,97){(a)}
\end{overpic}   
\begin{overpic}[scale=0.22]{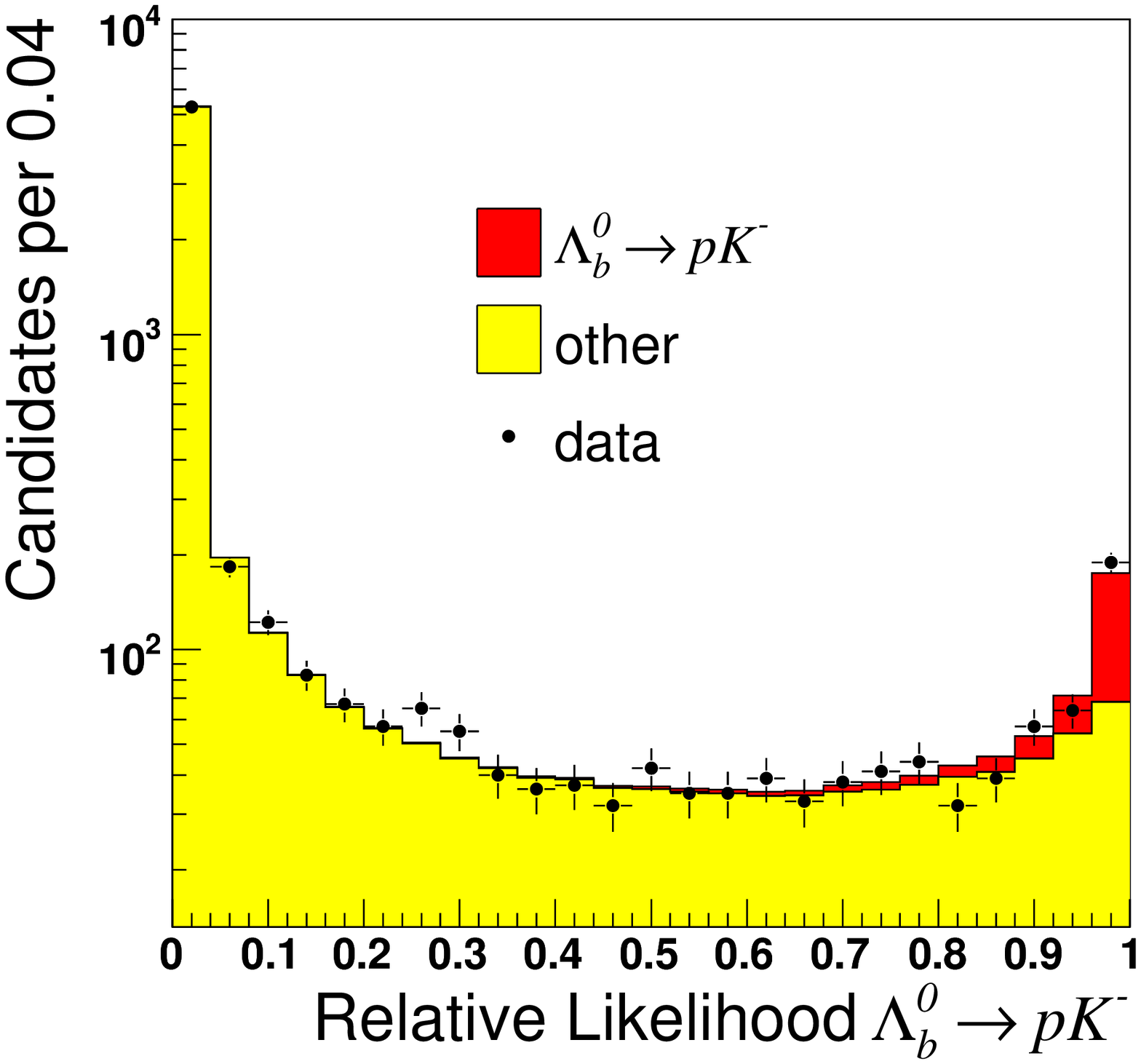}
\put(25,97){(b)}
\end{overpic}   
\begin{overpic}[scale=0.22]{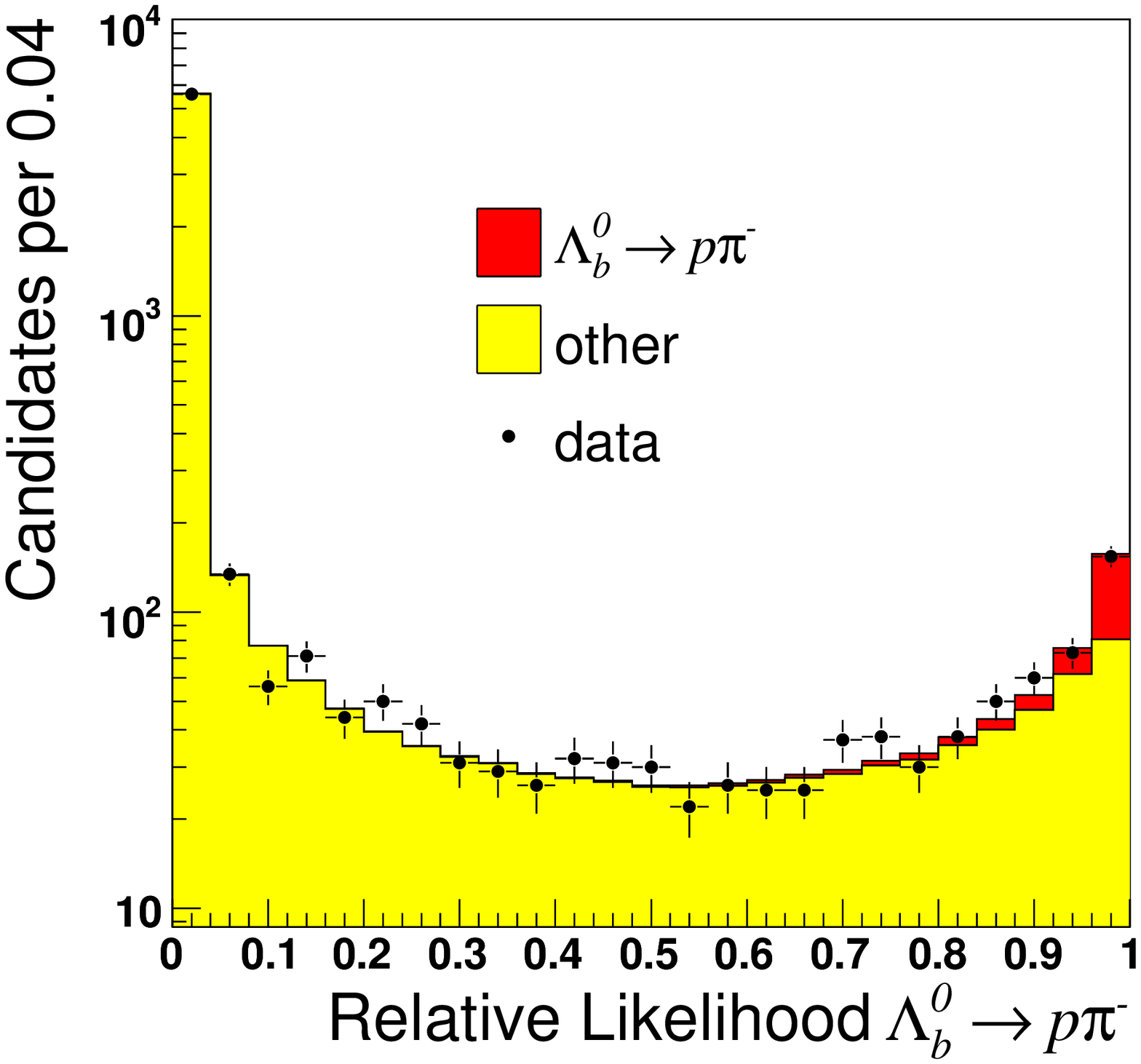}
\put(25,97){(c)}
\end{overpic}  
\caption{Distribution of the relative signal likelihood, $\like_{S}/(\like_{S}+\like_{{\rm other}})$, in the region 
$5.1 <\mpipi< 5.6~\massgev$. For each event, $\like_{S}$ is the 
likelihood for the \BsKpi\ (a), \LbpK\ (b), or \Lbppi\ (c) signal 
hypotheses, and $\like_{\rm other}$ is the likelihood for everything 
but the chosen signal, i.e. the weighted combination of all other 
components according to their measured fractions. Points with error 
bars show the distributions of data and histograms show the distributions predicted from the measured fractions.}
\label{fig:LRplots}
\end{figure}

From the signal fractions returned by the likelihood fits we calculate the signal yields shown in Table~\ref{tab:yields}.
The significance of rare unobserved signals is evaluated as the ratio of the 
yield observed in data, and its total uncertainty (statistical and systematic)
as determined from a simulation where the size of that signal is set to zero. 
This evaluation assumes a Gaussian distribution of yield estimates, 
supported by the results obtained from repeated fits to simulated samples.
This procedure yields a more accurate measure of significance with 
respect to the purely statistical estimate obtained from $\sqrt{-2\Delta {\rm ln}({\cal L})}$.
Significant signals are seen for \Bdpipi, \BdKpi, and \BsKK, previously observed by CDF~\cite{Abulencia:2006psa}.
We obtain significant signals for the \BsKpi\ mode ($8.2\sigma$), 
and for the \Lbppi\ ($6.0\sigma$) and \LbpK\ ($11.5\sigma$) modes.
Figure~\ref{fig:LRplots} shows relative likelihood distributions for 
these modes.
No evidence is found for the modes \Bspipi\ or \BdKK , in agreement 
with expectations of significantly smaller branching fractions. 
\begin{table}
\caption{\label{tab:yields} Yields of signals returned from fits. For rare unobserved modes significance is quoted.
The first quoted uncertainty is statistical, the second is systematic.}
{
\begin{tabular}{lcc}
\hline
Mode 	& N$_{s}$         &  Significance \\
\hline
\BsKpi	& 230 $\pm$ 34 $\pm$ 16	& $8.2 \sigma$  		\\
\Bspipi	& 26 $\pm$  16 $\pm$ 14	& $<3\sigma$	\\
\BdKK	& 61 $\pm$  25 $\pm$ 35	& $<3\sigma$	\\
\LbpK	& 156 $\pm$ 20 $\pm$ 11	& $11.5\sigma$		\\
\Lbppi	& 110 $\pm$ 18 $\pm$ 16	& $6.0 \sigma$          \\
\hline
\end{tabular}
}
\end{table}

To avoid large uncertainties associated with production cross sections 
and absolute reconstruction efficiency, we measure all branching 
fractions relative to the \BdKpi\ mode.
Frequentist upper limits~\cite{F-C} at the 90\% C.L. are quoted for the unseen modes.
For the measurement of \Lb\ branching fractions, the additional requirement 
$p_{T}(\Lb) > 6~\pgev$ was applied to allow easy comparison with other 
\Lb\ measurements at the Tevatron, which are only available above this 
threshold~\cite{Abulencia:2006df,Aaltonen:2008eu}. 
This additional requirement lowers the \Lb\ yields by about 20\%.

\section{\label{sec:fitresults} Acceptance corrections and systematics }

To convert the yields returned from the fit into relative branching fractions measurements,
we applied corrections for efficiencies of trigger and
offline selection requirements for different decay modes.
The relative efficiency corrections between modes do not exceed $8\%$
for the measurements of $b$--mesons and $40\%$ for \Lb\ branching fractions.
Most corrections were determined from the detailed
detector simulation, with some exceptions which were measured using data.
A momentum-averaged relative isolation efficiency between \Bs\
and \Bd\ of $1.00 \pm 0.03$  was determined from fully-reconstructed samples of
\Bs $\to J/\psi\,\phi$
and \Bd $\to J/\psi\,K^{*0}$~\cite{my_thesis}.
The lower specific ionization of kaons with respect to pions in the
drift chamber is responsible for a
$\simeq 5$\% lower efficiency to reconstruct a kaon.
This effect was measured in a sample of $D^{+}\to K^{-}\pi^{+}\pi^{+}$
decays triggered on two tracks, using the unbiased third track~\cite{Acosta:2004ts}.

The \Bspipi\ modes required a special treatment, since it
contains a superposition of the flavor eigenstates of the \Bs .
Their time evolution might differ from the one of the flavor-specific
modes if the width difference $\Delta\Gamma_{s}$
between the \Bs\ mass eigenstates is significant.
The current result was derived under the assumption that both modes are
dominated by the short-lived \Bs\ component, that $\Gamma_s=\Gamma_d$, and $\Delta\Gamma_s/\Gamma_s =
0.12\pm 0.06$~\cite{Beneke:1998sy,Lenz:2004nx}. 
 The latter uncertainty is included in estimating the overall systematic
uncertainty.

The dominant contributions to the systematic uncertainty are
the uncertainty on the combinatorial background model and
the uncertainty on the \dedx\ calibration and parameterization. 
Smaller systematic uncertainties are assigned for trigger efficiencies, physics background shape, kinematics,
$B$ meson masses and lifetimes.
\begin{table}[htb]
\caption{\label{tab:BR_summary} 
Measured relative branching fractions of
rare modes. The ratio $f_{\Lambda}/f_{d}$ is $p_T$--dependent~\cite{Aaltonen:2008eu}, and is 
defined here as: $f_{\Lambda}/f_{d} = \sigma(p\bar{p}\to\Lb X;p_{T} > 6~\pgev,|\eta|<1)/\sigma(p\bar{p}\to\Bd X;
p_{T} > 6~\pgev,|\eta|<1)$.
Absolute branching fractions were derived by normalizing to the current world--average value
${\mathcal B}(\mbox{\BdKpi}) = (19.4\pm 0.6) \times 10^{-6}$, and assuming the average values at high energy for the production fractions: 
$f_{s}/f_{d}= 0.276\pm 0.034$, and $f_{\Lambda}/f_{d} = 0.230 \pm 0.052$ \cite{PDG08}.
The first quoted uncertainty is statistical, the second is
systematic.}
{
\begin{tabular}{lcrclcc}
\hline
Mode && \multicolumn{3}{c}{Relative \BR}   && Absolute \BR ($10^{-6}$)\\
\hline
\BsKpi	&& \BsKpisuBdKpidef    &=&       0.071 $\pm$ 0.010 $\pm$ 0.007      && 5.0 $\pm$ 0.7 $\pm$ 0.8   \\
\Bspipi	&& \BspipisuBdKpidef   &=&       0.007 $\pm$ 0.004 $\pm$ 0.005      && 0.49 $\pm$ 0.28 $\pm$ 0.36 \\
	&&                     & &                                          && ($<1.2$ at 90\% C.L.)      \\
\BdKK	&& \BdKKsuBdKpidef     &=&       0.020 $\pm$ 0.008 $\pm$ 0.006      && 0.39 $\pm$ 0.16 $\pm$ 0.12 \\
	&&                     & &                                          && ($<0.7$ at 90\% C.L.)   \\
\LbpK	&& \LbpKsuBdKpidef     &=&       0.066 $\pm$ 0.009 $\pm$ 0.008      && 5.6 $\pm$ 0.8 $\pm$ 1.5   \\
\Lbppi	&&  \LbppisuBdKpidef   &=&       0.042 $\pm$ 0.007 $\pm$ 0.006      && 3.5 $\pm$ 0.6 $\pm$ 0.9      \\
\hline
\end{tabular}
}
\end{table}

\section{\label{sec:results}Results}

The final results on branching fractions are listed in Table~\ref{tab:BR_summary},
where $f_{d}$, $f_{s}$ and $f_{\Lambda}$ indicate the production fractions respectively of \Bd, \Bs\ and \Lb\ 
from fragmentation of a $b$ quark in $p\bar{p}$ collisions.
An upper limit is also quoted for modes in which no significant signal is
observed~\cite{F-C}. We also list absolute results obtained by normalizing the data to 
the world-average of \BR(\BdKpi)~\cite{PDG08,Aubert:2006fha}. The contributions
from the likelihood fit for each decay mode are shown in Fig.~\ref{fig:projections}.

The \CP--averaged branching fraction of the newly observed mode \BsKpi\ is consistent
with the previous upper  limit ($< 5.6 \times 10^{-6}$ at 90\%~C.L.)
based on a subsample of the current data~\cite{Abulencia:2006psa}, and 
agrees with the prediction in Ref.~\cite{bhh_scet}, 
but it is lower than most other predictions  \cite{th:QCDf,Sun:2002rn,Chiang:2008vc}.

The \Bspipi\ upper limit improves and supersedes the previous best 
limit~\cite{Abulencia:2006psa}.
The present measurement of $\BR(\BdKK)$ is in agreement with other existing measurements and has a 
similar resolution~\cite{PDG08},
but the resulting upper limit is weaker due to the observed central value.
The sensitivity to both \BdKK\ and \Bspipi\ is now 
close to the upper end of the theoretically expected 
range~\cite{th:QCDf,bhh_pQCD,bskpi_yu,Sun:2002rn,Li:2004ep}.  

We also report the first branching fraction measurements of charmless 
$\Lambda_b$ decays. They are
significantly lower than the previous upper limit of $2.3 \times 
10^{-5}$~\cite{Acosta:2005ab}, and
in reasonable agreement with predictions~\cite{Mohanta:2000nk}, thus excluding the possibility of
large ($O(10^{2})$) enhancements from R-parity violating 
supersymmetric scenarios~\cite{Mohanta:2000za}. 
Their ratio can be determined directly from our data with greater 
accuracy than the individual values. For this purpose, the 
additional $p_{T} > 6~\pgev$ requirement is not necessary, and we can exploit 
the full sample size, obtaining
$\BR(\Lbppi)/\BR(\LbpK) =  0.66 \pm 0.14 \pm 0.08$, 
in good agreement with the predicted range 0.60--0.62~\cite{Mohanta:2000nk}, but in
disagreement with the recent prediction in Ref.~\cite{Wei:2009np}.

The dominant systematic uncertainties of all measurements presented here are due to finite size of control samples 
and are expected to reduce with future extensions of the measurements.

In summary, we have searched for rare charmless decay modes of neutral
$b$--hadrons into pairs of charged hadrons in CDF data. 
We report the first observation of the modes \BsKpi, \Lbppi, and \LbpK,
and measure their relative branching fractions.
We set upper limits on the unobserved modes \BdKK\ and \Bspipi.

%


\end{document}

%% file: macro.tex
\newcommand{\comment}[1]{}

\def \rightdownarrow
 {\kern.3em
 \rule[.5ex]{.15mm}{2ex}
 {\mbox{$\kern-0.1em{\longrightarrow}$}}
 }
\def\lessim{\mathrel {\vcenter {\baselineskip 0pt \kern 0pt
\hbox{$<$} \kern 0pt \hbox{$\sim$} }}}
\def\gessim{\mathrel {\vcenter {\baselineskip 0pt \kern 0pt
\hbox{$>$} \kern 0pt \hbox{$\sim$} }}}
\newcommand{\beq}{\begin{equation}}
\newcommand{\eeq}{\end{equation}}
\newcommand{\bear}{\begin{array}}
\newcommand{\ear}{\end{array}}
\newcommand{\bet}{\begin{tabular}}
\newcommand{\eet}{\end{tabular}}
\newcommand{\beqn}{\begin{eqnarray}}
\newcommand{\eeqn}{\end{eqnarray}}

\newcommand{\bfh}{\begin{figure}[h]}
\newcommand{\efh}{\end{figure}[h]}

\newcommand{\tev}{\ensuremath{\mathrm{Te\kern -0.1em V}}}
\newcommand{\gev}{\ensuremath{\mathrm{Ge\kern -0.1em V}}}	
\newcommand{\mev}{\ensuremath{\mathrm{Me\kern -0.1em V}}}	
\newcommand{\kev}{\ensuremath{\mathrm{ke\kern -0.1em V}}}	
\newcommand{\massgev}{\mbox{\gev/$c^2$}}			
\newcommand{\massmev}{\mbox{\mev/$c^2$}}			
\newcommand{\pgev}{\mbox{\gev/$c$}}				

\newcommand{\stat}{\ensuremath{\mathit{~(stat.)}}}		
\newcommand{\syst}{\ensuremath{\mathit{~(syst.)}}}		

\newcommand{\CP}{CP}						
					
\newcommand{\bd}{\ensuremath{B^{0}}}				
\newcommand{\bs}{\ensuremath{B^{0}_s}}				

\newcommand{\bn}{\ensuremath{B^{0}_{(s)}}}			

\newcommand{\bhh}{\ensuremath{\bn \to h^{+}h^{'-}}}

\newcommand{\bshh}{\ensuremath{\bs \to h^{+}h^{'-}}}

\newcommand{\ppbar}{$\bar{p}p$}
\newcommand{\Bd}{\ensuremath{B^{0}}}

\newcommand{\Bu}{\ensuremath{B^{+}}}
\newcommand{\Bs}{\ensuremath{B_{s}^{0}}}
\newcommand{\Lb}{\ensuremath{\Lambda_{b}^{0}}}

\newcommand{\Bhh}{\ensuremath{\bn \to h^{+}h^{'-}}}
\newcommand{\Bdhh}{\ensuremath{\Bd \to h^{+}h^{'-}}}
\newcommand{\Bshh}{\ensuremath{\Bs \to h^{+}h^{'-}}}
\newcommand{\Lbph}{\ensuremath{\Lambda^{0}_{b} \to ph^{-}}}

\newcommand{\Bdpipi}{\ensuremath{\bd \to \pi^+ \pi^-}}

\newcommand{\BdKpi}{\ensuremath{\bd \to K^+ \pi^-}}

\newcommand{\BsKpi}{\ensuremath{\bs \to K^- \pi^+}}

\newcommand{\BsKK}{\ensuremath{\bs \to  K^+ K^-}}

\newcommand{\Bspipi}{\ensuremath{\bs \to  \pi^+ \pi^-}}

\newcommand{\BdKK}{\ensuremath{\bd \to  K^+ K^-}}

\newcommand{\Lbppi}{\ensuremath{\Lambda_{b}^{0} \to p\pi^{-}}}

\newcommand{\LbpK}{\ensuremath{\Lambda_{b}^{0} \to pK^{-}}}

\newcommand{\Dkpi}{\ensuremath{D^{0} \to K^- \pi^+}}

\newcommand{\DKpi}{\ensuremath{D^{0} \to K^- \pi^+}}

\newcommand{\dedx}{\ensuremath{\rm{dE/dx}}}
\newcommand{\like}{\ensuremath{\mathcal{L}}}

\newcommand{\BR}{\ensuremath{\mathcal B}}
\newcommand{\cdf}{CDF Collaboration}
\newcommand{\babar}{BaBar Collaboration}

\newcommand{\belle}{Belle Collaboration}

\newcommand{\cdfii}{CDF\,II}

\newcommand{\BsKpisuBdKpidef}{\ensuremath{\frac{\mathit{f_s}}{\mathit{f_d}}\frac{\BR(\BsKpi)}{\BR(\BdKpi)}}}
\newcommand{\BspipisuBdKpidef}{\ensuremath{\frac{\mathit{f_s}}{\mathit{f_d}}\frac{\BR(\Bspipi)}{\BR(\BdKpi)}}}
\newcommand{\BdKKsuBdKpidef}{\ensuremath{\frac{\BR(\BdKK)}{\BR(\BdKpi)}}}

\newcommand{\LbppisuBdKpidef}{\ensuremath{\frac{\mathit{f_{\Lambda}}}{\mathit{f_d}}\frac{\BR(\Lbppi)}{\BR(\BdKpi)}}}
\newcommand{\LbpKsuBdKpidef}{\ensuremath{\frac{\mathit{f_{\Lambda}}}{\mathit{f_d}}\frac{\BR(\LbpK)}{\BR(\BdKpi)}}}

\newcommand{\ie}{i.e.}					
 
\newcommand{\Lumi}{\ensuremath{\mathcal{L}}}			
\newcommand{\lumifb}{\mbox{fb$^{-1}$}}				

\newcommand{\mpipi}{\ensuremath{m_{\pi\pi}}}

%% file: bhh.bbl
\begin{thebibliography}{0}
\bibitem{Isolation}{Isolation is defined as
 $I_{B}=p_T(B)/(p_T(B)+\sum_ip_{Ti})$, where $p_T(B)$ is the transverse
 momentum of the $B$ candidate, and the sum runs over all other tracks within
 a cone of radius 1, in $\eta$-$\phi$ space around the $B$ flight-direction.}
\end{thebibliography}
